\documentclass[sn-mathphys,Numbered, Referee]{sn-jnl}
\usepackage{graphicx}%
\usepackage{multirow}%
\usepackage{amsmath,amssymb,amsfonts}%
\usepackage{amsthm}%
\usepackage{mathrsfs}%
\usepackage[title]{appendix}%
\usepackage{xcolor}%
\usepackage{textcomp}%
\usepackage{manyfoot}%
\usepackage{booktabs}%
\usepackage{algorithm}%
\usepackage{algorithmicx}%
\usepackage{algpseudocode}%
\usepackage{listings}%

\usepackage{cleveref}
\usepackage{dsfont}
\usepackage{mathtools}
\usepackage{nicefrac}
\usepackage{tabularx}

\newcommand{\R}{\mathbb{R}}
\newcommand{\Z}{\mathbb{Z}}
\renewcommand{\d}{{\textrm d}}

\theoremstyle{thmstyletwo}%
\newtheorem{remark}{Remark}%

\begin{document}
	
	\title[Phase transitions in 1D Riesz gases]{Phase transitions in one-dimensional Riesz gases with long-range interaction}
	
	\author*{\fnm{Rodrigue} \sur{Lelotte}\email{lelotte@ceremade.dauphine.fr}}
	
	\affil*{\orgdiv{CEREMADE}, \orgname{Université Paris Dauphine -- PSL}, \orgaddress{\street{Pl. du Maréchal de Lattre Tassigny}, \city{Paris}, \postcode{75016}, \country{France}}}
	
	\abstract{We provide numerical evidence for the existence of phase transitions with respect to the temperature in the one-dimensional Riesz gases with non-singular pair interaction, that is particles on the line interacting \emph{via} the potential $-|r|^{-s}$, where $s \in (-1, 0)$. Our numerics hint for the existence of two distinct phase transitions whose critical temperatures depend on $s$, namely a first transition which separates between a fluid and a quasisolid phase reminiscent of the Berezinski--Kosterlitz--Thouless (BKT) transition, and a second transition below which freezing occurs and the system is in a solid phase. We determine the phase diagram with respect to $s$ and the temperature $T$, which we find to be consistent with the known (or expected) results on the 1D Coulomb gas ($s = -1$), known to be a solid at all temperature, and the Dyson log--gas ($s = 0$) which exhibits a BKT transition at $T = 1/2$ and which is believed to be a fluid at all positive temperature. }
	

	\maketitle
	
	\section{Introduction}\label{sec1}
	It is well-known from the celebrated theorem of Hohenberg--Mermin--Wagner \cite{hohenberg_existence_1967, mermin_absence_1966, mermin_absence_1967, mermin_crystalline_1968}, as well as from an earlier result due to Van Hove \cite{van_hove_sur_1950}, that in one and two space-dimensions continuous symmetries can never be spontaneously broken at finite temperature, because long-range correlations are destroyed by thermal fluctuations. Nevertheless, a crucial assumption to these results is the short-range nature of the pair interaction. When the interaction remains strong at long distance, long-range fluctuations can persist in the thermodynamic limit, thus allowing for the existence of phase transitions. A seminal example was given by Dyson \cite{dyson_existence_1969} who proved the existence of a phase transition in the 1D \emph{Ising Ferromagnet} with low-decaying interactions between spins. Overall, many examples of 1D statistical lattice models where some breaking of symmetry occurs at finite temperature are known \cite{kittel_phase_1969, chui_pinning_1981, groskinsky_condensation_2003, sarkanych_exact_2017,  saryal_multiple_2018, cuesta_general_2004}. However, similar examples in the continuum seem much more scarcer in the literature. The 1D \emph{Coulomb gas}, which is a remarkable and completely integrable model, is an example of a one-dimensional system of classical particles in the continuum for which the translational-symmetry is broken at all temperature \cite{kunz_one-dimensional_1974, aizenman_structure_1980} — see also \cite{brascamp_inequalities_2002, jansen_wigner_2014} in the quantum case. One can also mention the works \cite{johansson_separation_1991,johansson_separation_1995} — although the nature of the transition is different than that of the one here sought.
	
\begin{figure}
		\centerline{\includegraphics[width=.65\textwidth]{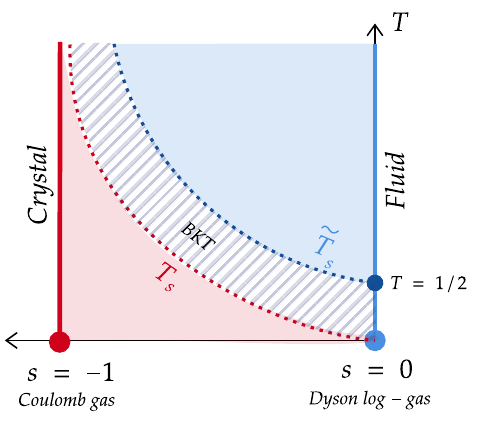}}
		\caption{Schematic phase diagram of 1D Riesz gases with respect to the temperature $T$ and the exponent $s$ of the interaction, following the intuition of \cite{lewin_coulomb_2022}. The 1D Riesz gas exhibits a phase transition at finite temperature $\tilde{T}_s>0$ separating between a fluid phase (blue area) and an ordered phase (red and hatched areas). At low enough temperature $T\ll \tilde{T}_s$, the system displays crystalline order (red area), whereas in the regime $0 \ll T < \tilde{T}_s$ we suspect a quasi-ordered phase of a BKT type. Therefore, we believe in the existence of two sets of critical temperatures, namely $\tilde{T}_s$ which separates between a fluid and a quasi-solid phase, and $T_s$ below which the Riesz gas is a solid — \emph{i.e.} crystal. The limiting behaviour of the critical temperatures with respect to $s$ are consistent with the phase diagrams of the Coulomb gas and the Dyson log-gas, corresponding to $s = -1$ and $s = 0$ respectively.}\label{fig:recapitulation}
	\end{figure}

	In this paper, we consider 1D \emph{Riesz gases}, that is particles on the line interacting through the pair potential $v_s(r) = \pm |r|^{-s}$, together with a uniform neutralizing background in the spirit of Jellium \cite{lewin_coulomb_2022}. We focus on the non-singular case, that is where the exponent $s$ ranges within $(-1, 0)$, in which case the sign of the interaction is chosen negative so as to make $v_s$ a repulsive potential. We remark that in this case, the potential does not decay at infinity. We provide numerical evidence which presumably rules in favor of the existence of a phase transition with respect to the temperature occurring at finite temperature. At high temperature, we found the pair correlation $g(r)$ to converge monotonically to the average density $\rho$ at large distance as in a fluid, whereas $g(r)$ displays long-lasting oscillations in the low temperature regime, accounting for the existence of a long-range order. At low enough temperature, we found the system to display crystalline features. From a closer investigation, we were led to suspect the existence of two distinct critical temperatures, hereafter denoted $\widetilde{T}_s$ and $T_s$. The first one separates between a fluid and a quasisolid phase reminiscent of the \emph{Berezinski--Kosterlitz--Thouless} (BKT) transition \cite{kosterlitz_ordering_1973, herbut_modern_2007, mudry_lecture_2014}, and another one below which the system is a true solid — see \Cref{fig:recapitulation}. 
	
	One-dimensional systems, despite their apparent oversimplicative physical traits, have been a continuously renewed source of exciting physics \cite{bernasconi_physics_2012, lieb_mathetical_1967}. Such models are usually more accessible to analytical calculations while being able to describe to a certain extent many problems of actual physical relevance. As for the 1D Riesz gases, they are interesting as they can be seen as the most natural interpolation familly between two important integrable models, both of which having received great shares of interest in physics and mathematics. Indeed, in the Coulomb case $s = -1$, one recovers as mentioned above the \emph{1D Jellium}, which is also called \emph{Coulomb gas} or 1D \emph{One-Component Plasma} (1dOCP). This is a beautiful and solvable statistical model which has been rather thoroughly investigated in the literature \cite{aizenman_structure_1980, aizenman_symmetry_2010,chafai_at_2020, choquard_statistical_1975, kunz_one-dimensional_1974}. In particular, the 1D Jellium is known to be crystallized at all temperatures \cite{kunz_one-dimensional_1974, aizenman_structure_1980}. On the other hand, in the limit $s \to 0^-$, by considering the first-order term of the pair interaction $v_s(r)$, we recover the \emph{Dyson log-gas} \cite{dyson_statistical_1962, dyson_statistical_1962-1, dyson_statistical_1962-2, dyson_statistical_1963}, that is particles on the line interacting \emph{via} the logarithmic interaction $-\ln|r|$. This model is of particular importance and regularly occurs in different areas of physics and mathematics. In the context of \emph{random matrix theory}, it is referred to as the \emph{$\beta-$ensemble} or \emph{sine-$\beta$ process} \cite{forrester_log-gases_2010, valko_continuum_2009-4}. For the special values $\beta = 1, 2$ and $4$, one recovers respectively the GOE (Gaussian Orthogonal Ensemble), GUE (Unitary, \emph{mutatis mutandis}), and GSE (Symplectic, \emph{idem}) ensembles \cite{forrester_log-gases_2010}. The log-gas is interesting from a statistical physics standpoint as an integrable toy model of particles interacting through a long-range and singular potential. We refer to the rather extensive \cite[Sec. V.C]{lewin_coulomb_2022} and the references therein, as well as the monograph \cite{forrester_log-gases_2010} for a very detailed account on the matter. As for its expected phase diagram, the Dyson log-gas is known to be crystallized at zero temperature \cite{sandier_1d_2015}, and it is believed that translation-invariance can never be broken at finite temperature \cite{requardt_wigner_1990}. A rigorous proof of this statement is given in \cite{erbar_one-dimensional_2021} in the case of stationary point processes \cite{dereudre_introduction_2019}, thus accounting for the case of the thermodynamic limt of the log-gas on the circle, that is with periodic boundary conditions.  Our work is then motivated by a question asked in the recent review \cite{lewin_coulomb_2022}, where it is wondered whether or not there exists a smooth transition curve between those two limiting cases, namely the Coulomb gas $s = -1$ and the Dyson log-gas $s = 0$. Our findings confirm this prediction.  
	
		\vspace{1em}

	\begin{remark}[BKT transition for $s > 0$]
		In this paper, we only investigate the case of negative exponents $s \in (-1, 0)$. For $s > 0$, it is expected that the translational symmetry will never be broken \cite{lewin_coulomb_2022}. This is known rigorously for $s > 2$ \cite{papangelou_absence_1987}. Nevertheless, it might be that the BKT transition — which is not associated to a broken symmetry — that appears for the Dyson log-gas at $T = 1/2$ (see \Cref{ssec:dyson} below) and which, according to our results, also exists for $s < 0$ (see \Cref{fig:recapitulation}) does not cease to exist for $s > 0$, at least up to some threshold value of $s$. It would be interesting to investigate this question.
	\end{remark}

	\section{Riesz, Coulomb and Dyson gases}\label{sec:def}
	In this section, we define the periodic Riesz gases in one space-dimension. We discuss the special cases of the Coulomb gas and of the Dyson log-gas. As a sanity-check of our algorithm, which we will use later to study the Riesz gas with general exponent $-1 < s < 0$, we present numerics on the log-gas which are seen to be consistent with known — or at least suspected — theoretical results. These numerics might be of independent interest to some readers. 
	
	\subsection{Definition of the periodic 1D Riesz gases}
	In the long-range case $s < 1$, the periodic 1D Riesz gas is defined as follows. We consider $N$ particles constrained to the segment $\ell_L = [0, L]$ and we impose periodic boundary conditions to supress possible boundary effects in our numerical experiments. In the spirit of Jellium \cite{lewin_coulomb_2022}, we add a compensating uniform background of opposite charge with density $\rho = \nicefrac{N}{L}$ to ensure charge neutrality whence summability. In the periodic setting, this amounts to deleting the Fourier zero mode of the interaction potential, see below \eqref{def:v_per_fourier}. A key parameter in the study of phase transitions is $\Gamma^{-1} := \rho^{-s} T$, where $T$ is the effective temperature of the system. By scaling we will suppose without loss of generality that $\rho = 1$, otherwise stated that $N = L$, so that our parameter of interest is the sole effective temperature $T$. The associated periodic Riesz potential $\widetilde{v}_{s, L}$ can be analytically expressed using special functions \cite{lewin_coulomb_2022} (see \Cref{rem:per_pot} below) contrary to higher dimensions where one needs to ressort to some numerical computations often relying on \emph{Ewald summation}. This potential is defined by its Fourier transform \cite[Sec. IV.A.2]{lewin_coulomb_2022} up to here unimportant multiplicative constant as 
	\begin{equation}\label{def:v_per_fourier}
		\widehat{\widetilde{v}_{s, L}}(k) = \sum_{\substack{k \in 2 \pi \Z/L \\ k \neq 0}} \frac{\delta_k}{|k|^{1-s}}.
	\end{equation}
	The 1D Riesz gas is then formally defined as the system obtained in the thermodynamic limit, that is by considering the large $N$ limit of the canonical ensemble $Q_N$ defined as the Gibbs measure with density
	\begin{equation}\label{eq:can_ens}
		Q_N(r_1, \dots, r_N) = \frac{1}{Z(s, \beta, N)} \exp \left(- \beta\sum_{1 \leq i < j \leq N} \widetilde{v}_{s, N}(r_i - r_j)\right),
	\end{equation}
	where $Z(s, \beta, N)$ is the usual \emph{partition function}, that is the normalizing constant such that $Q_N$ is a probability measure on the $N$-torus. Here $\beta$ is the inverse temperature, that is $\beta = \nicefrac{1}{T}$.  We can define the \emph{canonical free energy} of the 1D Riesz gas as the thermodynamic limit
	\begin{align}\label{eq:free_energy}
		f(s, \beta) := \lim_{N \to \infty}  \frac{-\beta^{-1}\log Z(s, \beta, N)}{N}.
	\end{align}
	The existence of this limit at all temperature was proved in \cite{lewin_coulomb_2022} extending an argument of \cite{hardin_next_2015}. At zero temperature the energy per unit length is exactly known, and the system is crystallized \cite{borodachov_discrete_2019}. We also mention the works of Serfaty \emph{et al.} \cite{serfaty_coulomb_2015, lewin_coulomb_2022} where the cases $s < 0$ are not treated \emph{sic} but are covered by the theory to some extent\footnote{S. Serfaty. \emph{Personal communication}}. To the best of our knowledge, no other theoretical results are rigorously known except for those mentioned above. In particular, the convergence of the correlation functions in the thermodynamic limit seems to be unknown for $-1 < s < 0$ at the present time. We recall that the \emph{$k$-point correlation function} $\rho^{(k)}(r_1, \dots, r_k)$ is defined as
	\begin{equation}\label{eq:correlations_fn}
		\rho^{(k)}(r_1, \dots, r_k) = \frac{N!}{(N-k)!} \int_{\R^{N-k}} Q_N(r_1, \dots, r_k, r_{k+1}', \dots, r_N') \d r_{k+1}' \dots \d {r_N}'
	\end{equation}
	While the correlation functions are very important in the study of phase transitions, they are also very useful from a mathematical standpoint as they completely characterise the limiting object obtained from the canonical ensemble $Q_N$ as one considers the thermodynamic limit  $N \to \infty$. This limiting object is a (\emph{Gibbs}) \emph{point process} \cite{lewin_coulomb_2022,dereudre_introduction_2019}. We emphasize that the question of its existence and \emph{casu quo} of its uniqueness — related to the (non-)existence of phase transitions — while well-studied in the short-range case $s > d$ in any dimension $d$, see  \cite{ruelle_statistical_1999, georgii_canonical_1976, georgii_gibbs_2011, doerushin_existence_1967}, is a complicated and subtle problem which remains mainly open in the long-range case $s < d$, see \cite[Sec. III]{lewin_coulomb_2022}. In dimension $d = 1$, it was only very recently studied by Dereudre and Vasseur \cite{dereudre_number-rigidity_2023} and Boursier \cite{boursier_decay_2022} in the case $0 < s < 1$. In the logarithmic case $s = 0$, it is studied in \cite{dereudre_dlr_2021-1}. To the best of our knowledge, the case of negative exponents $s < 0$ seems to have been eluded in the literature so far, at the exception of the Coulomb case $s = -1$, which has been extensively studied, see \emph{e.g.} \cite{kunz_one-dimensional_1974, choquard_statistical_1975, aizenman_structure_1980, lenard_exact_2004, aizenman_symmetry_2010, chafai_at_2020}.

	In this work, we will focus our attention on the two-point correlation function $\rho^{(2)}(r, r')$, which we called the \emph{pair correlation}. Here, the correlation between two particles only depends on their distance from one another, so that the pair correlation can be written as a function of a single variable hereafter denoted $$g(r) := \rho^{(2)}(0,r).$$ The function $g(r)$ describes how the density of particles varies as a function of distance from a given reference particle. In the case of a crystal, $g(r)$ is a periodic function with sharp maxima at the lattice site. On the other hand, in the case of a perfect fluid such as an ideal gas, the particles are independent of each other, so that $g(r)$ is constant. More generally, in the absence of long-range order, the density fluctuations between two particles should decrease rapidly at large distances, that is $g(r)$ should converge rapidly to the average density $\rho$, whereas in the presence of long-range order, $g(r)$ should display a slower decay and/or oscillations at large $r$. 
	
	We will also investigate the (\emph{static}) \emph{structure factor} $S(k)$ \cite[Chap. 4]{hansen_theory_2013}, see also \cite{dinnebier_powder_2008,gingrich_structure_2004, sirota_complete_1989, yarnell_structure_1973}. The structure factor is defined in the thermodynamic limit $N \to \infty$ as the Fourier transform of the truncated pair correlation $g(r) - 1$, namely 
	\begin{equation}\label{eq:def_S(k)}
		S(k) := 1 + \frac{1}{2\pi}\int_{\R} e^{-irk} (g(r) - 1) \d r.
	\end{equation}
	In the finite length $N < \infty$, the above definition should be modified accordingly by considering the (discrete) Fourier transform on the circle $L^2(\R /N \Z)$, in which case $S(k)$ in only defined on $k \in \Z/N$. If the pair correlation $g(r)$ is oscillating, the structure factor should have a peak at $k = 1$ or more generally at any multiple of the period of $g(r)$ — so called \emph{Bragg peak} in condensed matter physics and crystallography. On the other hand, if the pair correlation rapidly converges to the average density of our system, as expected in a fluid phase, the structure factor should be a smooth function of the wavenumber $k$. 
	
	\begin{figure}
		\centerline{\includegraphics[width=\textwidth]{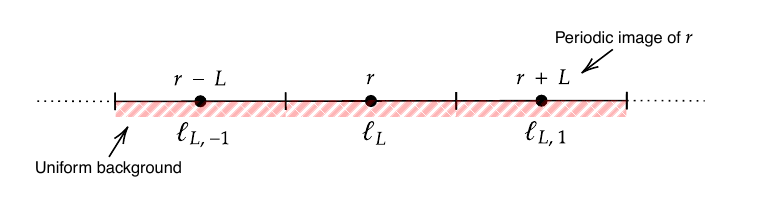}}
		\caption{The periodic Riesz potential $\widetilde{v}_{s, L}$ is obtained by considering that each particles located at $r \in \ell_L$ interacts with its periodic images in the $\ell_{L,k}$ and the uniform background.} \label{fig/periodic_potential}
	\end{figure}
	
	\vspace{1em}
	
	\begin{remark}[Periodic Riesz potentials]\label{rem:per_pot}
		The periodic Riesz potential $\widetilde{v}_{s, L}$ on the line is defined, as for any periodization of a general potential, as the sum of the interactions between a particle located at $r \in \ell_L$ and its periodic images in the segments $r_k \in \ell_{L, k}$ that is $r_k := r + kL$ for all $k \in \Z$ (see \Cref{fig/periodic_potential}). Therefore, we have 
		\begin{align}
			\widetilde{v}_{s, L}(r) = \sum_{k \in \Z} v_s(r + Lk).
		\end{align}
		We remark that in the short-range case $s > 1$, this infinite sum is convergent and is closely related to the \emph{Hurwitz zeta function} $\zeta(s, r)$, as it can be expressed as
		\begin{equation}\label{eq/per_riesz_pot}
			\widetilde{v}_{s, L}(r) = \zeta(s, r) + \zeta(s, 1-r)  \quad \text{when } s > 1.
		\end{equation}
		In the long-range case $s < 1$ the resulting series is evidently divergent. To ensure summability — at least when $s > -1$ — one may add in the spirit of Jellium a uniform background of opposite charge over each $\ell_{L, k}$ in such a way as to ensure charge neutrality of the overall system. The periodic potential is then expressed for all $s > -1$ as
		\begin{equation}\label{eq/def_per_pot}
			\widetilde{v}_{s, L}(r) = \lim_{q \to \infty} \left( \sum_{|k| \leq q} v_s(r + k L) - \rho \int_{ \cup_{|k| \leq q} \ell_{L, k}} v_s(r - r') \d r' \right).
		\end{equation} 
		It turns out rather beautifully that this normalization, which we emphasize to be very natural from the viewpoint of physics, exactly corresponds to the meromorphic extension to the half complex plane $\{ \Re(s) > -1\}$ of the periodic Riesz potential in the short-range case $s > 1$ with a pole at $s = 1$ \cite{borwein_convergence_1985, borwein_energy_1988, borwein_analysis_1989, borwein_lattice_2014, lewin_coulomb_2022}. Therefore, $\widetilde{v}_{s, L}$ rewrites as in \eqref{eq/per_riesz_pot} if one agrees to use the meromorphic continuation of the Hurwitz zeta function to the punctured complex plane $\mathbb{C} \setminus \{1\}$ on the right-hand side. We note that, although this entails that the periodic potential $\widetilde{v}_{s, L}$ can actually continuated over the entire complex plane — at the exception of $s = 1$ — one should be aware that the above formula \eqref{eq/def_per_pot} is \emph{a priori} only valid when $\Re(s) > - 1$. Pushing $s$ below this threshold usually requires another kind of normalization. We refer to \cite[Section IV. A]{lewin_coulomb_2022} for further details on this question and more generally on the analytic continuation of the periodic Riesz potential in arbitrary dimension $d \geq 1$. 
	\end{remark}
	
	\subsection{The 1D Coulomb Gas}
	
	In this section, we review important results regarding the \emph{1D Coulomb gas}, also known as the \emph{1D Jellium} or 1D \emph{One-Component Plasma} (1dOCP). This remarkable model, which corresponds to the choice $s = -1$, was extensively studied by Kunz in \cite{kunz_one-dimensional_1974}, where the thermodynamic limit of the free energy \eqref{eq:free_energy} and the correlations functions $\rho^{(k)}(r_1, \dots, r_k)$ — see \eqref{eq:correlations_fn} — are computed through transfer matrix techniques \cite{baxter_exactly_2016}. It is to be noted that the Coulomb case is very special, as the force between two distinct particles does not depend on their mutual distance, so that ordering the particles on the line somehow leads to a form of ``conditional independence''. This fortuitous property was leveraged by Aizenman and Martin in \cite{aizenman_structure_1980} where similar results as of \cite{kunz_one-dimensional_1974} are proved using the electric field as the key variable and appealing to ergodic arguments to conclude. Directly extending theses methods to other values of $s$ seems complicated, if not impossible. Altogether, the authors of \cite{kunz_one-dimensional_1974, aizenman_structure_1980} managed to prove that the correlation functions were proper periodic functions at all temperature, and thus that the Coulomb gas is crystallized at all temperature. We shall now briefly explain the strategy of both papers, and verify the results numerically.
	
	\begin{figure}
		\centerline{\includegraphics[width=\textwidth]{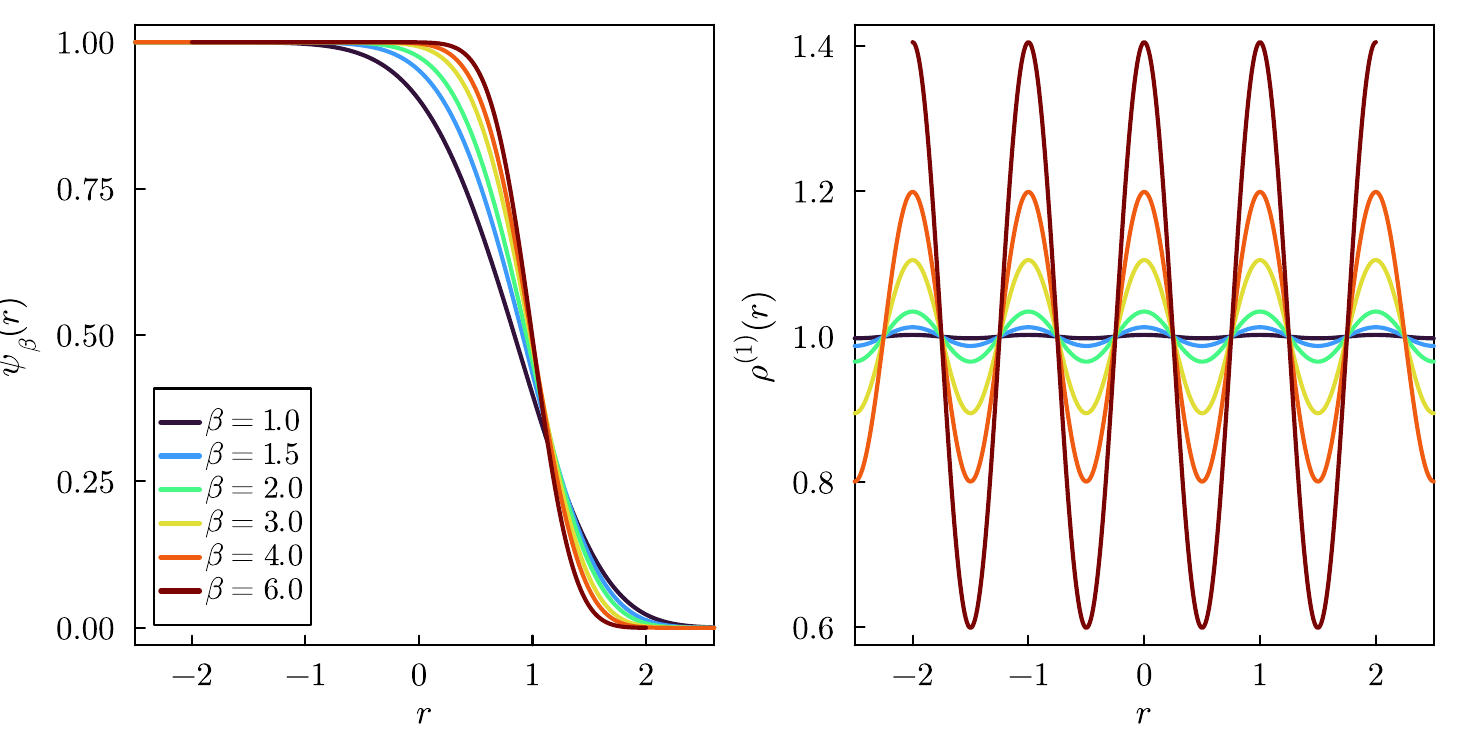}}
		\caption{We compute the Perron--Frobenius eigenvector $\psi_\beta$ of the operator $K$ defined in \eqref{eq:def_K} by a straightforward discretization for several temperatures. We then compute the density $\smash{\rho^{(1)}(r)}$ using \Cref{eq:density_kunz}. We observe that the density is a proper periodic function of period $\rho = 1$ as proved by Kunz. We also retrieve the properties that $\psi_\beta$ converges to $1$ as $r \to -\infty$ and to $0$ as $r \to \infty$, and that $\psi_\beta$ converges to the Heaviside function centered at $r = 1$ in the vanishing temperature limit, see \cite[Appendix \& p. 315]{kunz_one-dimensional_1974}. All figures in this work were made using \texttt{Julia} and the \texttt{Plots.jl} package. } \label{fig:kunz_1D}
	\end{figure}
	
	The strategy of Kunz in \cite{kunz_one-dimensional_1974} essentially boils down to the celebrated \emph{transfer-matrix method} in statistical physics \cite{baxter_exactly_2016}. The strategy is used by Kunz both with free and periodic boundary conditions. It should be noted that the argument heavily relies on both the one-dimensional nature of the system and on the very peculiar form of the Coulomb potential in dimension $d =1$, namely $-|r|$. Indeed, the Jellium energy — that is, when the particles interact with each other as well as with the uniform compensating background, see \eqref{eq:jel_1d} below — is a quadratic function once restricted to the set of ordered configuration. Indeed, if we suppose that $-N/2 \leq r_1 \leq \dots \leq r_N \leq N/2$, the Jellium energy in the non-periodic setting rewrites as
	\begin{multline}\label{eq:jel_1d}
		-\sum_{1 \leq i<j \leq N} |r_i - r_j| + \sum_{i = 1}^N\int_{-\frac N2}^{\frac N2} |r_i -r| \d r - \frac12 \int_{-\frac N2}^{\frac N2} \int_{-\frac N2}^{\frac N2} |r - r'| \d r \d r' \\
		= \sum_{i = 1}^N \left(r_i - i + \frac{N + 1}{2}\right)^2 + \frac{N}{2}.
	\end{multline}
	In particular, the canonical Gibbs measure $Q_N$ associated to the above energy is a Gaussian once restricted to the set of ordered configurations. Using this property, Kunz was able to rewrite the free energy $f_N(\beta)$ in the finite length $N$ and at inverse temperature $\beta$ as
	\begin{equation}
		f_N(\beta) = \left\langle g_\beta, K^N g_\beta\right\rangle_{L^2(\R_+)}.
	\end{equation}
	Here, $K$ is a compact operator with positive kernel over the Hilbert space $L^2(\R_+)$ which serves as an infinite-dimensional analogous to the so-called \emph{transfer matrix} and $g_\beta$ is an explicit function in $L^2(\R_+)$. We emphasize that $K$ depend on the inverse temperature $\beta$ but does \emph{not} depend on the number of particles $N$. It is given by the operator
	\begin{equation}\label{eq:def_K}
		K f(r) := \int_{r-1}^\infty e^{-\beta u^2} f(u) \d u,
	\end{equation}
	which is an integrable operator with kernel $K(r, r') = e^{-\beta r'^2} \mathds{1}(r' \geq r- 1 )$. Appealing to the \emph{Perron--Frobenius theorem} \cite[][Lem. 1 in Appendix]{kunz_one-dimensional_1974} it follows from positivity and compactness of $K$ that it has a simple largest eigenvalue $\lambda(\beta)$ associated to an unique positive normalized eigenfunction $\psi_\beta \in L^2(\R_+)$. By discretizing the operator $K$ as defined in \eqref{eq:def_K} above, we can compute numerically the eigenvector $\psi_\beta$, see \Cref{fig:kunz_1D}. The thermodynamic limit of the free energy can then be readily expressed using those quantities \cite[Eq. (17)]{kunz_one-dimensional_1974}. The correlation functions can be dealt with in a very similar manner. For instance, Kunz found that the one-point correlation function $\rho^{(1)}(r)$ converges to the periodic function given by 
	\begin{equation}\label{eq:density_kunz}
		\rho^{(1)}_\tau(r)  = \sum_{k \in \Z} \psi_\beta(-r - k - \tau) \psi_\beta(r + k + \tau) 
	\end{equation}
	for some $\tau \in \R$ \cite[Eq. (40-41)]{kunz_one-dimensional_1974}. In fact, the scalar $\tau$ depends on the sequence of the number of particles $N$'s considered in the thermodynamic limit, which is very clear manifestation of the breaking of symmetry. Furthermore,  Kunz managed to prove that all the correlation functions were periodic. Nevertheless, to ensure that crystallization really happens, one still needs to prove that those functions are properly periodic, that is that they are not constant function. By appealing to analyticity \cite[][at p. 314]{kunz_one-dimensional_1974}, he managed to prove this fact at low enough temperature $\beta \gg 1$. This was eventually generalized to all temperature $\beta > 0$ by Aizenman and Martin \cite{aizenman_structure_1980}. In \Cref{fig:kunz_1D}, we show $\rho^{(1)}_\tau(r)$ computed for several temperatures. Finally, Kunz studied the problem with periodic boundary conditions, in which case he found that the correlation functions $\smash{\rho^{(k)}_{\rm per}}$ were all obtained by averaging their counterparts $\rho^{(k)}_\tau$ in the non-periodic setting over their period, that is 
	\begin{equation}\label{eq:corr_fn_per}
		\rho^{(k)}_{\rm per}(r_1, \dots, r_k) = \int_0^1 \rho^{(k)}_\tau(r_1, \dots, r_k) \d \tau.
	\end{equation}
 It follows from \eqref{eq:corr_fn_per} that crystallization cannot be detected on  $\rho^{(1)}(r)$ anymore in the periodic setting, and that one should look at the pair correlation $g(r)$, which will then be a proper periodic function of the distance $r$. This is a clear manifestation of the breaking of symmetry. In what will follow regarding the Riesz gas with general exponent $- 1 < s < 0$, we will look at the pair correlation for the same reasons.

\begin{figure}
		\centerline{\includegraphics[width=\textwidth]{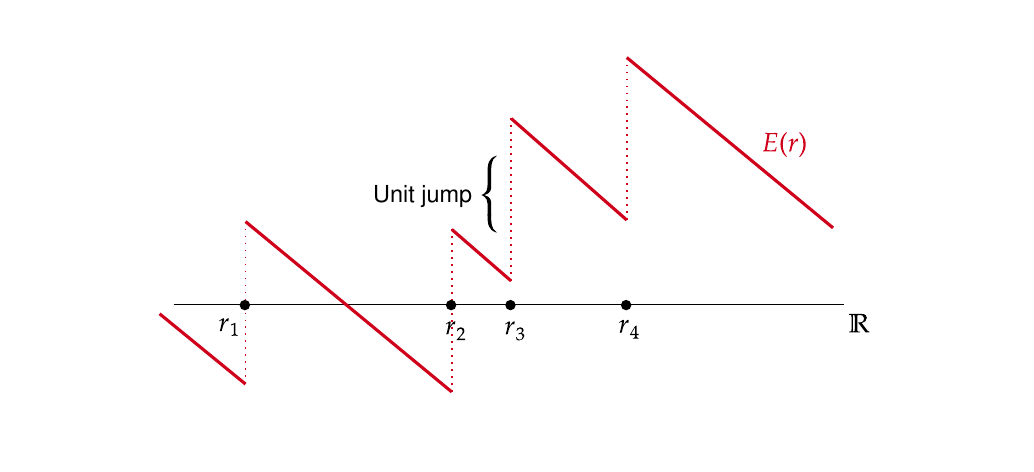}}
		\caption{The electric field generated by a configuration of particles $r_1, \dots, r_N$ has a very nice structure in the case of the Coulomb potential. It is a piecewise linear function with unit slope and jump of unit size located at each particle $r_j$. The position of the particles can therefore be retrieved from the positions of the jumps. } \label{fig/electric_field_coulomb}
	\end{figure}

	Aizenman and Martin \cite{aizenman_structure_1980}, building on previous works of Lenard \cite{lenard_exact_2004, lenard_exact_2004-1} and Edwards--Lenard \cite{edwards_exact_2004}, took a different road than that of Kunz to study the 1D Coulomb gas. Their key idea is to work with the \emph{electric field} $E(r)$ rather than the particles themselves. Indeed, there is a one-to-one correspondence between the set of configurations and the set of possible electric fields, as proved in \cite[Lem. 4]{aizenman_structure_1980}. This can be seen from the fact that, given a configuration of particles $X = (r_1, \dots, r_N)$, the electric field $E_X(r)$ generated by this configuration has a very simple structure, as it is a piecewise linear function of unit slope with a jump of unit size located at each particle $r_j$, see \Cref{fig/electric_field_coulomb}. This allows to view the electric field as a random jump process whose semigroup can be readily expressed \cite[Eq. (49) sqq.]{aizenman_structure_1980}. The most important thing to stress out here is the \emph{Markovian} nature of this process, as can be intuited from \Cref{fig/electric_field_coulomb}. This allows yet once more to appeal to Perron--Frobenius to cope with the thermodynamic limit, see \cite[Eq. (4.11)]{aizenman_structure_1980}. Finally, using an ergodic theorem, Aizenman and Martin proved the periodicity of the correlations functions for all $\beta > 0$, thus extending the result of \cite{kunz_one-dimensional_1974} mentioned earlier. They were able to show that the associated limiting point process obtained in the thermodynamic limit $N \to \infty$ can be defined by the usual set of characterisations such as \emph{Dobrushin--Lanford--Ruelle} (DLR),  \emph{Bogoliubov–Born–Green–Kirkwood–Yvon} (BBGKY) and \emph{Kubo-Martin-Schwinger} (KMS) equations — we refer the reader to \cite{lewin_coulomb_2022, dereudre_introduction_2019} on this matter.                                                                           
	
	We note that the electric field is a very convenient variable in the case of the Coulomb potential because the energy \eqref{eq:jel_1d} can be expressed as a positive quadratic form in the variable $E_X$ using the \emph{carré du champ} operation \cite[Eq. (2.7) and (2.9)]{aizenman_structure_1980}. For arbitrary exponent $s$, this very useful \emph{carré du champ} is no longer available \emph{as-is}. Nevertheless, it is still possible  — although much more involved —  to work with the electric field rather that the particles themselves. We refer to the long line of work initiated by Serfaty and collaborators — see \cite{serfaty_coulomb_2015} for a self-contained reference, or the references in \cite{lewin_coulomb_2022}.
	
	\subsection{The Dyson log--gas}\label{ssec:dyson}
	
	The \emph{Dyson log-gas} \cite{dyson_statistical_1962, dyson_statistical_1962-1, dyson_statistical_1962-2, dyson_statistical_1963}, which corresponds to the choice $s = 0$ and for which we recover the logarithmic interaction $v_0(r) = - \ln(r)$, is also accessible to analytical computations for specific values of the inverse temperature $\beta$, using the tenacious analogy between log-gases and random matrix models \cite{forrester_log-gases_2010}. We recall that, in the context of random matrices, the Dyson log-gas is called the $\beta$-ensemble. Using a fairly general theory of one-dimensional quantum fluids of Haldane \cite{haldane_effective_1981}, it was conjectured by Forrester in \cite{forrester_analogues_1984} that the leading term in the expansion of the pair correlation $g(r)$ at large $r$ would be given by
	\begin{equation}\label{eq:g_log}
		g(r) \underset{r \to \infty}{\sim} \begin{dcases}
			1 - \frac{1}{\pi^2 \beta r^2} \quad & \text{for }\,\, \beta < 2,\\
			1 + \frac{\cos (2\pi r)}{2 \pi^2 r^2} - \frac{1}{2 \pi^2 r^2} \quad & \text{for }\,\, \beta = 2,\\
			1 + c \frac{\cos 2\pi r}{r^{4/\beta}} - \frac{1}{2 \pi^2 r^2}\quad &\text{for }\,\, \beta > 2
		\end{dcases}
	\end{equation}
	for some universal constant $c > 0$. The expansion is rigorous for $\beta = 1, 2$ and $4$ as can be shown using the analogy between Dyson log-gas and standard Gaussian ensembles \cite{forrester_log-gases_2010}. It is also veraricous in the case of even or rational $\beta$'s as proved in \cite{forrester_exact_1993} and \cite[Chap. 13]{forrester_log-gases_2010}.  We see that the decay of $g(r)$ in the large $r$ limit exhibits a transition at $\beta = 2$ from an universal monotonous power-law decay $r^{-2}$ to an oscillating and non-universal decay whose power depends on the temperature. This is a celebrated example of a  \emph{Berezinski--Kosterlitz--Thouless} (BKT) transition \cite{kosterlitz_ordering_1973}, see also \cite{herbut_modern_2007, mudry_lecture_2014}. In the vanishing temperature limit $\beta \to \infty$, as the oscillations become predominant, $g(r)$ converges to a periodic function and the system is crystallized onto a (floating) \emph{Wigner crystal} \cite{sandier_1d_2015, leble_uniqueness_2015}.

	It follows from the expansion \eqref{eq:g_log} that the behaviour of the structure factor $S(k)$ \eqref{eq:def_S(k)} in the small wavenumber limit $k \to 0$ is to be given by that of the $-1/r^2$ term to the leading order. Indeed, although the leading term of the (truncated) pair correlation is of order $1/r^{4/\beta}$ as soon as $\beta > 2$, the cosine term shifts its contribution to the Fourier transform at $k \sim 1$. More generally, all the oscillating terms $\cos(2\pi n r)/r^{4n/\beta}$ in the expansion of the pair correlation of the log-gas when $\beta > 2$ — which we did not write in \eqref{eq:g_log}, see \cite{forrester_analogues_1984} —  only contribute to the structure factor at $k \sim n$. Altogether, the term $-1/r^2$ is the only one which contributes to the behaviour of $S(k)$ near $k = 0$, so that
	\begin{equation}\label{eq:S_bkt_0}
		S(k) \sim 2 \beta^{-1} |k| \quad \text{ as } k \to 0.
	\end{equation}
	It also follows from the expansion \eqref{eq:g_log} that $S(k)$ should feature a singularity at $k = 1$ as soon as $\beta \geq 4$. This singularity will be logarithmic at the threshold $\beta = 4$ and should diverge as an inverse power-law when $\beta > 4$, that is (up to multiplicative constant)
	\begin{equation}\label{eq:S_bkt_1}
		S(k) \sim \frac{1}{|1 - k|^{1 - 4\beta^{-1}}} \quad \text{ as } k \to 1^-.
	\end{equation}
	In any case, we emphasize that this singularity is of an integrable type. This is in clear constrast with what one would expect in a crystal. Indeed, as explained earlier, in the case of a crystal the structure factor $S(k)$ should have a sharp peak at $k = 1$ corresponding to that of a Dirac mass, as expected from the periodic nature of the pair correlation $g(r)$.

	\begin{figure}
		\centerline{\includegraphics[width=\textwidth]{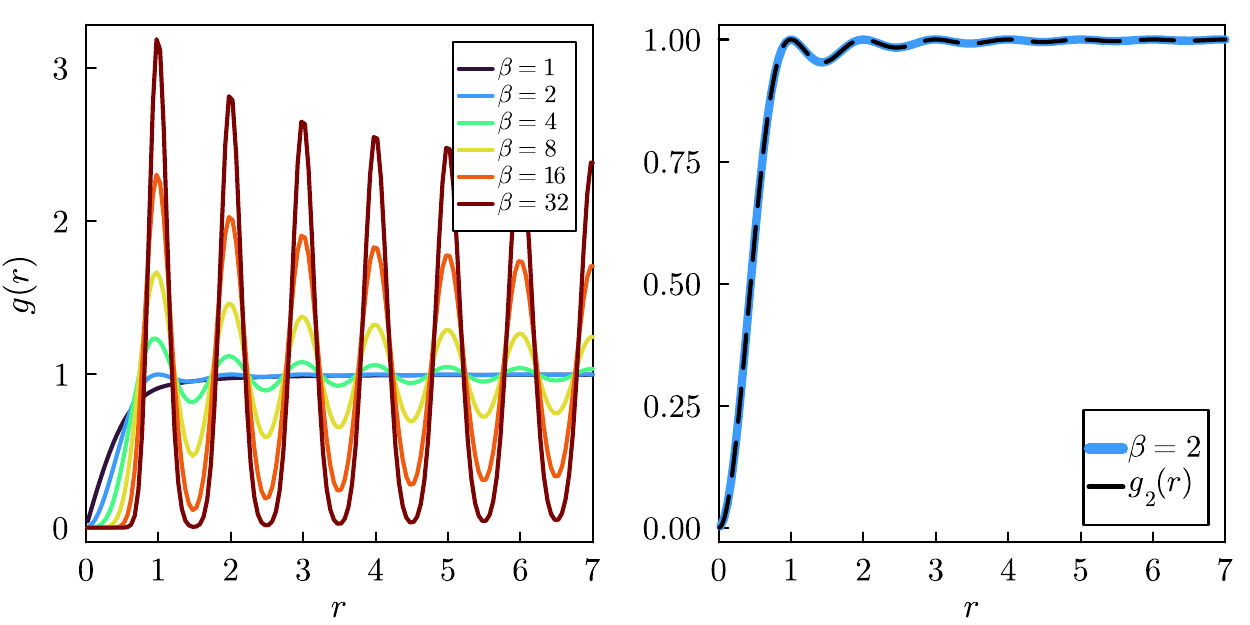}}
		\caption{On the left, we display the pair correlation $g(r)$ for the Dyson log-gas at several temperatures obtained by our algorithm. Below the critical temperature $\beta = 2$, the pair correlation converges monotonically to the average density — here set to $\rho = 1$ — whereas above the critical temperature $g(r)$ we observe oscillations which eventually vanish as $g(r)$ converges to $\rho$ in the large $r$ limit. On the right, the approximation of $g(r)$ at $\beta = 2$ obtained numerically is seen to be consistent with the exact formula for the pair correlation \cite{forrester_log-gases_2010}. We used $N = 100$ particles and built the pair correlation by binning.}\label{fig:log_gas_pc}
	\end{figure}
	
	\begin{figure}
		\centerline{\includegraphics[width=\textwidth]{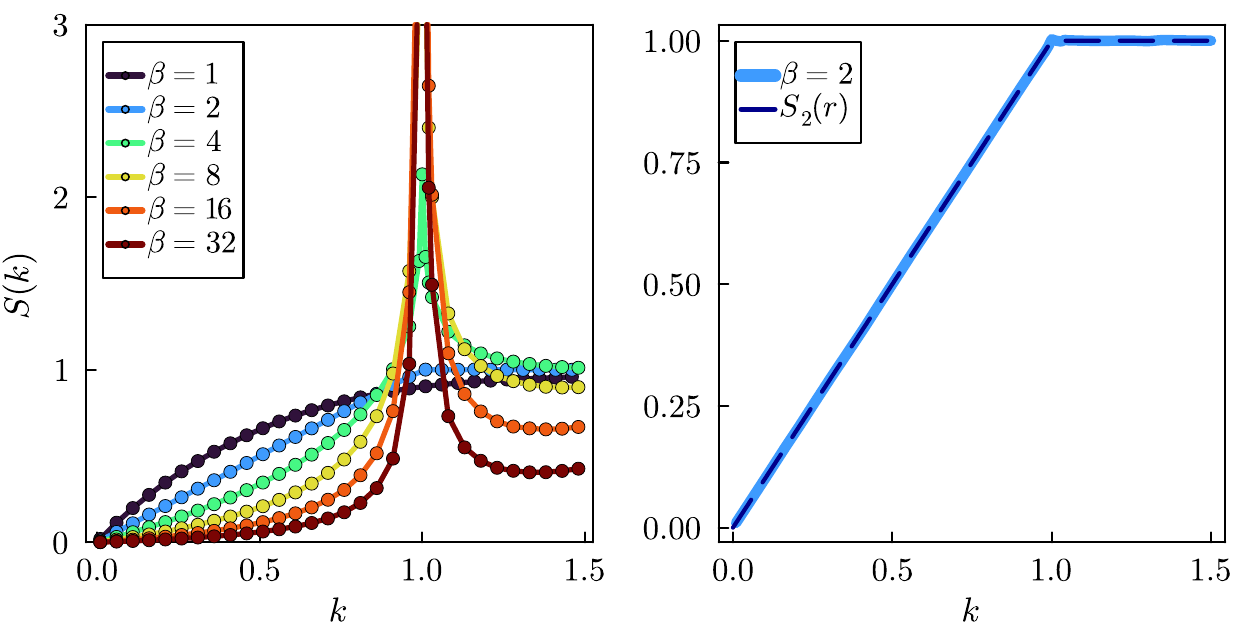}}
		\caption{On the left, we display the structure factor $S(k)$ for the Dyson log-gas at several temperatures, using the pair correlations previously computed and \eqref{eq:def_S(k)}. Above the critical temperature $\beta = 2$, the structure factor displays a peak at $k = 1$ as expected from the \eqref{eq:S_bkt_0}. On the right, the approximation of $S(k)$ at $\beta = 2$ obtained numerically is seen to be consistent with the exact formula for the pair correlation \cite{forrester_log-gases_2010}. We thinned the number of displayed wavenumbers for visual convenience. }\label{fig:log_gas_sf}
	\end{figure}
	
\begin{figure}[!ht]
		\centerline{\includegraphics[width=\textwidth]{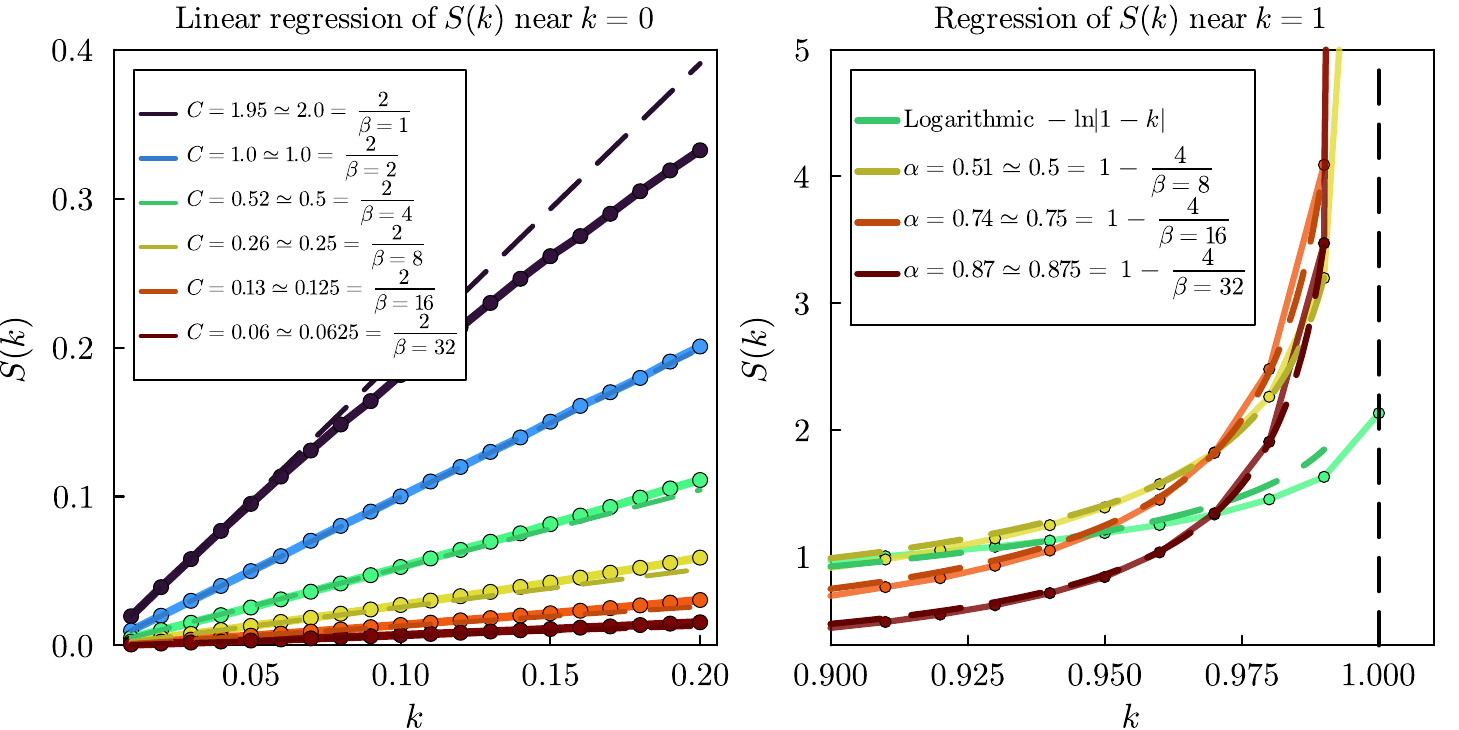}}
		\caption{On the left, we regress the structure factor $S(k)$ of the Dyson log-gas near $k = 0$ on $C|k|$. The slope coefficient $C$ is found to be close to $2 \beta^{-1}$, which is consistent with \eqref{eq:S_bkt_0} — and therefore consistent with the Forrester's expansion \eqref{eq:g_log}. On the right, we regress the structure factor $S(k)$ near $k = 1$ as $k \to 1^+$ on $c|1 - k|^{\alpha}$ for some here unimportant constant $c$. We found $\alpha$ to be closed to $1 - 4\beta$ for $\beta > 4$, which is consistent with \eqref{eq:S_bkt_0}. When $\beta = 4$, the structure factor behaves logarithmically $S(k) \sim \log|1-k|$, as expected.} \label{fig:log_gas_regress}
	\end{figure}

	Although our main goal is this paper is to investigate the long-range situation $-1 < s < 0$, we show that the previous claims on the behaviour of $S(k)$ near $k = 0$ \eqref{eq:S_bkt_0} and near $k = 1$ \eqref{eq:S_bkt_1} are confirmed numerically. Although we do so as a sanity-check for our algorithm, which is presented in \Cref{app}\footnote{Our code is available at \url{https://github.com/rodriguel/PTRiesz}.}, these results may be of independent interest for some readers.  In \Cref{fig:log_gas_pc}, we display an approximation of the pair correlation $g(r)$ obtained numerically for several inverse temperatures $\beta$. We observe that below the critical temperature $\beta = 2$, the correlation converges monotonically to the average density $\rho = 1$. From the critical temperature $\beta = 2$ further on, the pair correlation $g(r)$ displays damped oscillations whose amplitude strengthen as the temperature is further decreased, and eventually $g(r)$ converges to the average density in the large $r$ limit. Our approximation fits perfectly with the exact formula for $g(r)$ at $\beta = 2$ \cite{forrester_log-gases_2010}. In \Cref{fig:log_gas_sf}, we display the associated structure factor $S(k)$, which is obtained by computing the (discrete) Fourier transform of the pair correlation $g(r)$ as in \eqref{eq:def_S(k)}. We then regress $S(k)$ near $k = 0$ to obtain the universal behaviour of the slope at which $S(k)$ approaches $k = 0$, and we regress at $k = 1$ to obtain the exponent at which $S(k)$ diverges. The results, displayed in \Cref{fig:log_gas_regress}, are in clear adequation with \eqref{eq:S_bkt_0} and \eqref{eq:S_bkt_1} — and therefore are numerical confirmations of the veracity of the expansion \eqref{eq:g_log} conjectured by Forrester.

	\section{Evidence for the existence of a phase transition}\label{sec:results}
	
	In this section, we investigate numerically the 1D Riesz gases for general exponent $-1 < s < 0$. In \Cref{ssec:evidence}, we show evidence which strongly advocates for the existence of a phase transition with respect to the temperature $T$ depending on $s$. In \Cref{ssec:deter}, we further study the behaviour of the pair correlation $g(r)$ and most importantly of the structure factor $S(k)$ to make our claim that there coexists two separate phase transitions, namely a fluid--quasisolid transition of a BKT-type similar to that of the Dyson log-gas, and a freezing point below which the system is crystallized. A phase diagram is then determined numerically according to a set of criteria summarized in \Cref{tab:summary}.
	
	\subsection{Long-range order at low enough temperature and existence of a critical temperature}\label{ssec:evidence}
	
\begin{figure*}
		\centerline{\includegraphics[width=\textwidth]{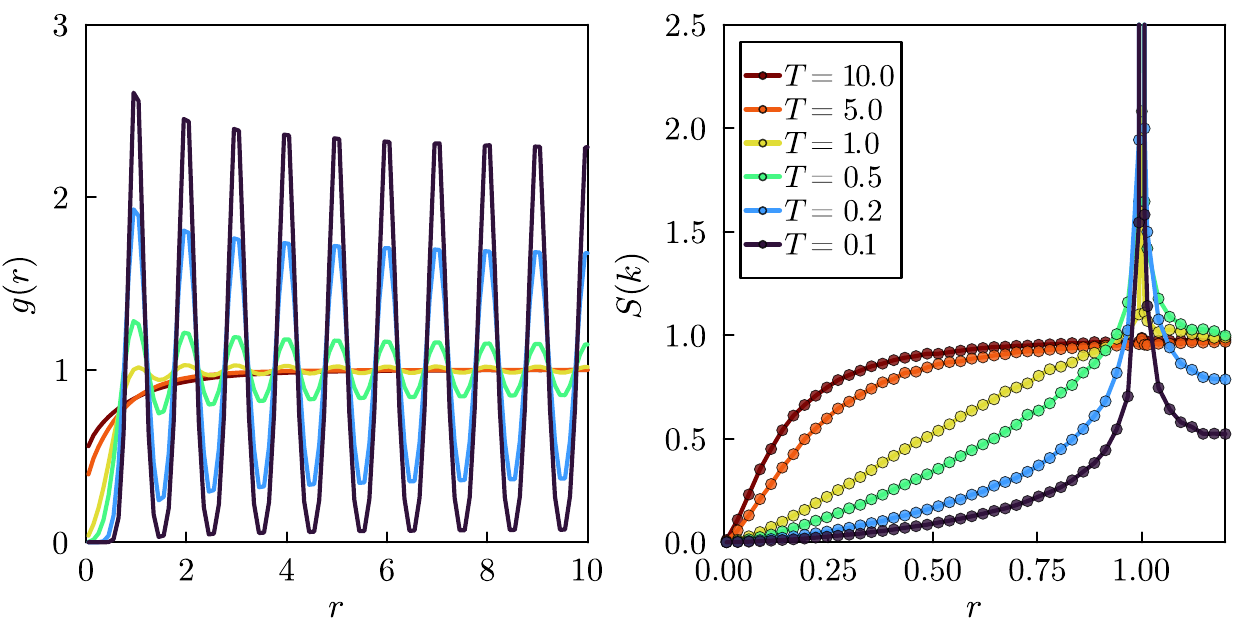}}
		\caption{On the left (resp. right) we display the pair correlation $g(r)$ (resp. the structure factor $S(k)$) at various temperatures for the Riesz gas of exponent $s = -0.5$. At high enough temperature, $g(r)$ converges rapidly and monotonically to the average density, whereas in the low temperature regime it features persistent oscillations whose amplitude increases as the temperature is furthered lowered. In this regime, we observe that the structure factor $S(k)$ has a sharp peak at $k = 1$  whose width (resp. height) decreases (resp. increases) as the temperature is lowered, hinting the presence of a Dirac mass accounting for the periodicity of $g(r)$ in the large $r$ limit. We used $N = 150$ particles. }\label{fig:hint_pt}
	\end{figure*}

\begin{figure*}[!ht]
		\centerline{\includegraphics[width=\textwidth]{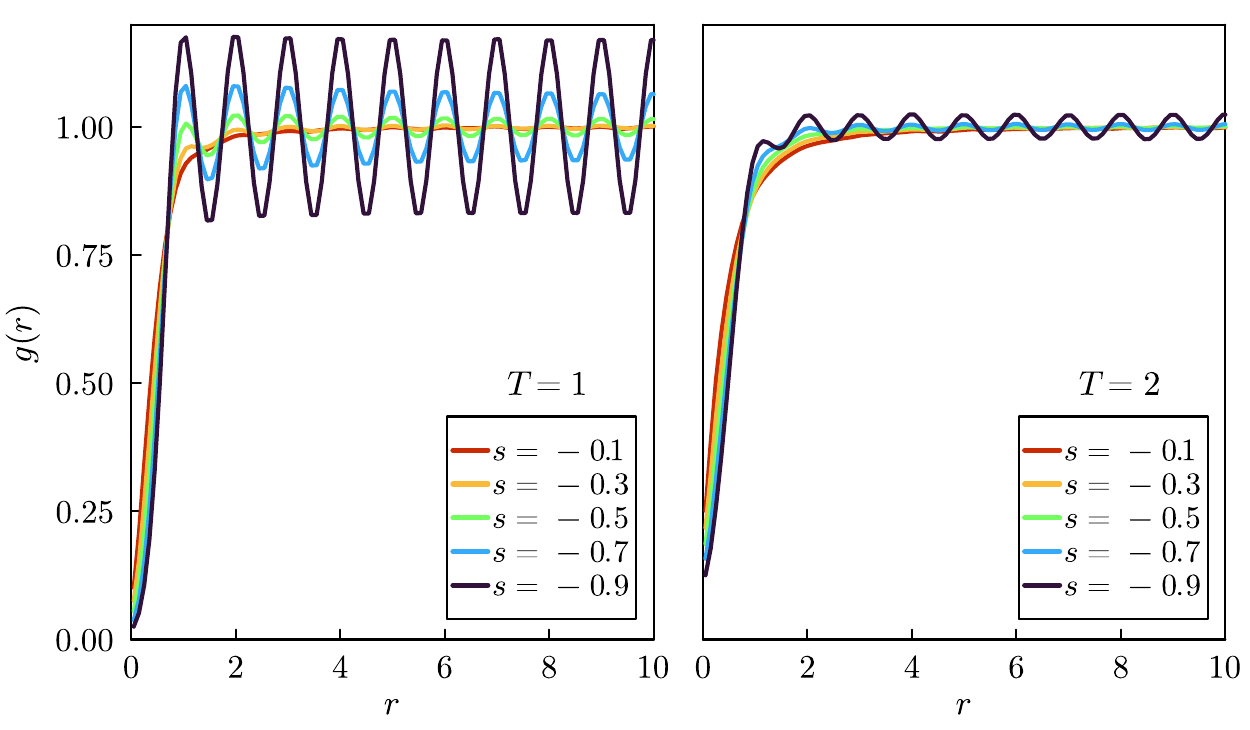}}
		\caption{We fix two temperatures $T = 1$ (left) and $T = 2$ (resp. right) and we vary the exponent $s$. We observe that the oscillations of $g(r)$ appear at lower temperatures as the exponent $s$ gets closer to the Coulomb gas $s = -1$, which is consistent with the fact that the Coulomb gas is crystallized at all temperature. On the other hand, the oscillations appear at much lower temperature as $s$ gets closer to the Dyson log-gas $s = 0$, which is consistent with the fact that the Dyson log-gas is expected to be a fluid at all positive temperature. From this, we are led to claim that the critical temperature $ \mathfrak{T}_s$ depends on the exponent $s$ in such a way as to interpolate between the phase diagrams of the 1D Coulomb gas and the Dyson log-gas, as conjectured in \cite{lewin_coulomb_2022}. }\label{fig:hint_pt_2}
	\end{figure*}

	We compute the pair correlation $g(r)$ at several temperatures for various exponents $-1 < s < 0$. As a first evidence for the (non-)existence of some long-range order, we wonder whether or not $g(r)$ features persistent oscillations at low enough temperature, or equivalently if the structure factor $S(k)$ has a sharp peak at $k = 1$. From our experiments, we see that for all $s$ in the range $-1 \leq s \leq 0$, at high enough temperature $g(r)$ rapidly and monotonically converges to the average density as $r \to \infty$, as one would expect in a fluid phase, whereas for small enough temperatures it exhibits long-lasting oscillations which are ever more amplified as the temperature is further lowered, consistent with the fact that the system is crystallized at zero temperature \cite{borodachov_discrete_2019}. This is clearly seen in \Cref{fig:hint_pt}, where we display $g(r)$ and $S(k)$ for the 1D Riesz gas for $s = -0.5$ at varying temperature. Furthermore, we found this conspicuous qualitative change of behaviour to occur within a range of temperatures depending on the exponent $s$. In \Cref{fig:hint_pt_2}, we fix two temperatures and we vary the exponent $s$. We observe that the oscillations appear sooner, that is at higher temperature, as the exponent $s$ gets closer to $s = -1$, and conversely that they appear later as $s$ gets closer to $s = 0$. This is consistent with the fact that the Coulomb gas is crystallized at all temperature and with the fact that the Dyson log-gas is expected to be a fluid at all positive temperature.
	
	From what precedes, we are brought to believe in the existence of a critical temperature $\mathfrak{T}_s$ depending on the exponent $-1 < s < 0$ which separates between a fluid phase in the high temperature regime $T > \mathfrak{T}_s$ and an ordered-phase in the low temperature regime $T < \mathfrak{T}_s$. The critical temperature should interpolate between the Coulomb gas, that is $\mathfrak{T}_s \to \infty$ as $s \to - 1$, and the Dyson log-gas, that is $ \mathfrak{T}_s \to 0$ as $s \to 0$. Nevertheless, a clear determination of $\mathfrak{T}_s$ is evidently complicated, pertaining to both the underlying limitations of numerics and the absence of an absolute criterion to either rule in favor or out of the appearance of a long-range order. Furthermore, it is unclear whether or not the oscillations which appear in the pair correlation $g(r)$ eventually vanish in the large $r$ limit, as in the Berezinski--Kosterlitz--Thouless paradigm. The rest of our paper is dedicated to obtain a better understanding of $\mathfrak{T}_s$.
	
	\subsection{Determination of the critical temperature and nature of the transition}\label{ssec:deter}
	It remains to determine the nature of the transition which was put in evidence in the previous section, as well as the behaviour of the transition curve with respect to $s$, which was loosely denoted $\mathfrak{T}_s$ above. In this section, we give several criteria to determine whether the system is in a fluid, quasisolid or solid phase. In fact, we make the claim that there actually coexists two distinct set of critical temperatures, denoted hereafter $\widetilde{T}_s$ and $T_s$. The first one separates between a fluid and a quasisolid phase reminiscent of the BKT transition and similar to that of the Dyson log-gas discussed earlier, while the second one corresponds to the point at which the system is frozen onto a true solid. We use these criteria to give a — at least schematic — phase diagram of the Riesz gas with respect to the effective temperature $T$ and the exponent $-1 < s < 0$. 
	\begin{figure*}[!ht]
		\centerline{\includegraphics[width=\textwidth]{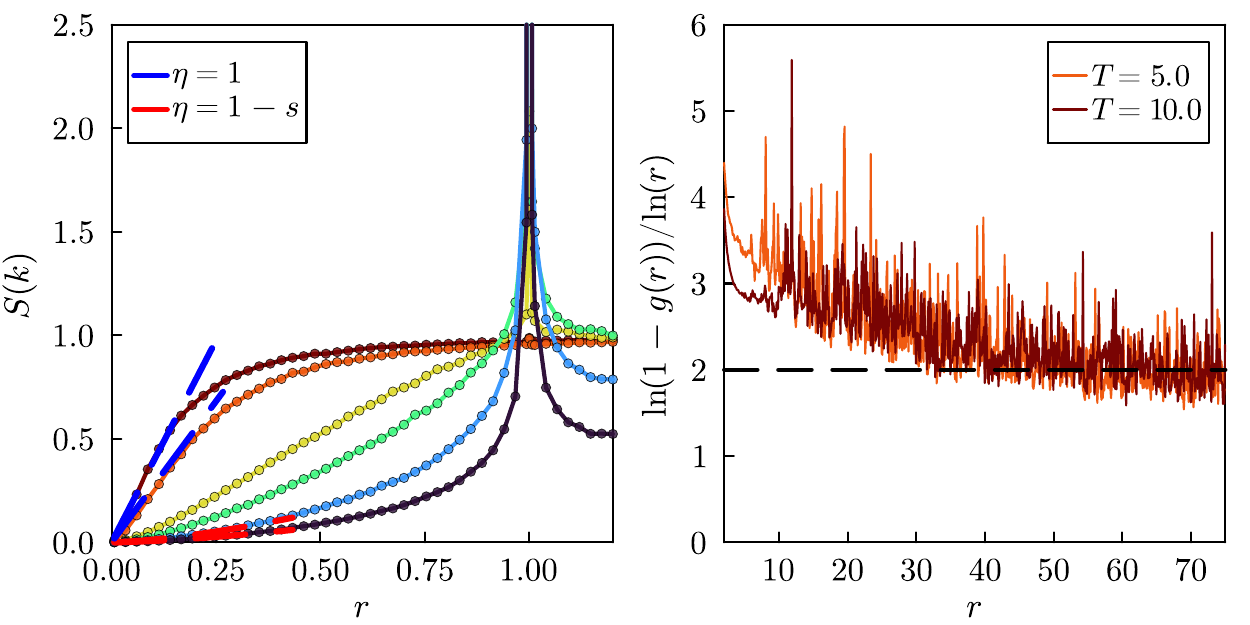}}
		\caption{On the left, we regress the structure factor $S(k)$ near $k = 0$ on $C|k|^\eta$, for the same values as of \Cref{fig:hint_pt}. We allow $\eta \in \{1, 1-s\}$, where $\eta = 1-s$ is the natural exponent for the crystallisation regime, whereas $\eta = 1$ corresponds to that of a liquid. In the high temperature regime, we see that $S(k)$ behaves linearly as $k \to 0$. This is coherent with the behaviour of the pair correlation at large $r$, which decays as $-1/r^2$, as seen on the second figure on the right, where we display $\ln(1 - g(r))/\ln(r)$, which is seen to approach the value $2$ as $r$ gets large.  At low enough temperature the choice $\eta = 1-s$ is found to yield a smaller regression residue, which is a strong evidence that the Riesz gas is crystallized according to Aizemann \emph{et al.} \cite{aizenman_bounded_2001}}\label{fig:deter_1}
	\end{figure*} 
	
	\subsubsection{Behavior of $S(k)$ in the limit $k\to0$}  
	
	In the preceding section, we were attentive to whether or not the pair correlation $g(r)$ converges to a periodic function in the large $r$ limit, as this is a clear manifestation of the crystallization. This is evidently related to the appearance in the Fourier space of a peak at $k = 1$, or for that matters at any multiple of the period of $g(r)$.  Nevertheless, it turns out that the breaking of symmetry can be seen on the behaviour of the structure factor in the limit $k \to 0$. We stress out that this is a non-trivial fact. Indeed, Aizemann, Goldstein and Lebowitz gave in \cite{aizenman_bounded_2001} a sufficient condition for translational symmetry to be broken in one-dimensional systems. This result, which is related to the notion of \emph{hyperuniformity} \cite{torquato_local_2003, torquato_hyperuniform_2018}, essentially says that if the structure factor $S(k)$ behaves like $|k|^\eta$ in the small wavenumber vicinity $k\to0$ for some $\eta > 1$, then translational symmetry must be broken in the thermodynamic limit. We note that this is \emph{not} in contradiction with the expected phase diagram of the Dyson log-gas, for which $\eta = 1$ as shown previously. 
	
	On the other hand, it follows from the extension of a heuristical argument of Forrester \cite{forrester_log-gases_2010} — see also \cite{alastuey_decay_1985} — that, if the Riesz gas at exponent $s$ is crystallized, then $S(k)$ must behave like $S(k) \simeq C|k|^{1-s}$ in the small wavenumber limit $|k| \to 0$ for some constant $C > 0$. The exponent $\eta = 1-s$ is very natural, as it fits with the Dyson log-gas, for which $\eta = 1$, and the Coulomb gas for which $\eta = 2$. It also fits with the results of obtained by Boursier \cite{boursier_decay_2022} in the case where $0 < s < 1$. The argument of Forrester, which can be found in \cite[Chap. 11]{forrester_log-gases_2010} in the case of the log-gas, can be extended as follows for any $s$. If we pertub our system at equilibrium by a fluctuating charge density $\epsilon e^{-ikr}$, and if we denote by $\rho_\epsilon(r)$ the density of the pertubed system, then it must be that 
	\begin{equation}\label{eq:forrester}
		\rho_\epsilon(r) - \rho(r) \sim_{k \to 0} - \epsilon e^{ikr} 
	\end{equation}
	where $\rho(r)$ is the density of the original system. We emphasize that this equivalence is only formal and \emph{a priori} not rigourous. It says that the system responds in an appropriate manner to the perturbation, that is in such a way as to cancel the perturbation and remain in equilibrium. In fact, this can be viewed as characteristic of a crystalline order. Indeed the crystal should be able to remain stable under perturbation of large enough wavelength $\lambda \gg a$ — or small enough wavenumber $k \ll 1/a$, as in \eqref{eq:forrester} — where $a$ is the crystal constant.
	
	Now, by letting $\epsilon \to 0$ in \eqref{eq:forrester}, and using the well-known relations which links the functional derivatives of the free energy and the correlation functions \cite{hansen_theory_2013}, the left-hand side of \eqref{eq:forrester} can be written as 
	\begin{equation}
		\rho_\epsilon(r) - \rho(r) \sim - \epsilon \beta \int_{\R} W_\epsilon(r') \rho^{(2)}(r, r')
	\end{equation}
	where $W(r) := - |r|^{-s} \ast e^{ikr}$ is the potential associated to the charge density which perturbs the system, and $\rho^{(2)}(r, r')$ is the two-point correlation as defined earlier \eqref{eq:correlations_fn}. By using the invariance by translation, we get
	\begin{equation}\label{eq:1-s_S}
		\rho_\epsilon(r) - \rho(r) \sim - \epsilon \beta \int_{\R} W_\epsilon(r') \rho^{(2)}(r, r') \sim \epsilon \beta \frac{1}{|k|^{1+s}} S(k).
	\end{equation}
	Putting \eqref{eq:forrester} and \eqref{eq:1-s_S} yields that $\eta = 1-s$. Furthermore, the coefficient $C$ should be linear in the temperature, and in fact it should be given by $C = 2 \beta^{-1}$ similarly to the Dyson log-gas, as seen in \eqref{eq:S_bkt_0}. In \Cref{fig:deter_1}, we observe that at high enough temperature, the structure factor $S(k)$ converges linearly to $0$ as $k \to \infty$. This is seen to be consistent with the truncated pair correlation $g(r) - 1$ decaying as $ - 1/r^2$ in the large $r$ limit. On the contrary, as the temperature is decreased, the structure factor $S(k)$ flattens near the origin, and at low enough temperature it is seen to decrease sublinearly to $0$ as $k \to 0$, in fact as $|k|^{1-s}$ as expected from the above heuristic.

	\subsubsection{Behaviour of $S(k)$ in the limit $k \to 1$}
	
	\begin{figure}
		\centerline{\includegraphics[width=\textwidth]{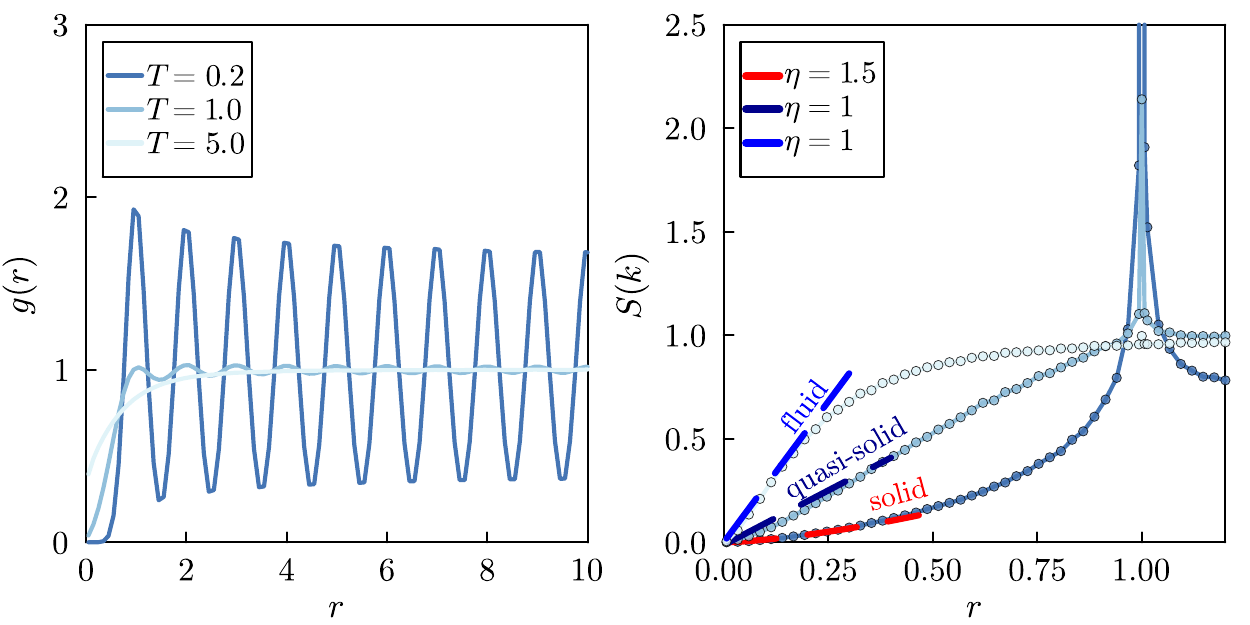}}
		\caption{From a closer investigation on the structure factor (right), we are led to believe in the existence of three distinct regimes for $s = -0.5$. At high temperature, the structure factor behaves linearly near the origin and it has no peak at $k = 1$, hinting a fluid phase. As the temperature is decreased, a peak appears at $k=1$, but the structure factor remains linear in the limit $k \to 0$. This is characteristic of a quasisolid. Finally, at low enough temperature, $S(k)$ flattens out near the origin and behaves as $|k|^{1-s}$ : it is a crystal. }\label{fig:deter_2}
	\end{figure}

	In the case of a crystal, the structure factor $S(k)$ should have a sharp peak at $ k =1$ corresponding to a Dirac mass. We recall that in the case of the Dyson log-gas, the structure factor also has a peak at $k=1$ for $\beta > 4$, but it is an integrable function and not a Dirac mass — and, in the case $2 < \beta < 4$, the function $S(k)$ is non-monotonous near $k = 1$. In \Cref{fig:deter_1}, we see that the appearance of the peak at $k = 1$ precedes that of the flattening of the structure factor near the origin, and that there exists a range of temperatures for which the peak exists but the structure factor seems to behave linearly near the origin. This is clearly seen for $ s= -0.5$ in \Cref{fig:deter_2}.

	We are therefore brought to believe in the existence of two distinct phase transitions. At high enough temperature, the Riesz gas is a fluid. As the temperature is decreased down to a certain threshold, there is a BKT transition similar to that of the Dyson log-gas as discussed earlier, corresponding to the formation of a quasisolid. Eventually, as the temperature is furthered decreased, there is another threshold at which the system is frozen into a true crystal. This is clearly depicted in \Cref{fig:deter_2}.
	
	\begin{figure*}
		\centerline{\includegraphics[width=\textwidth]{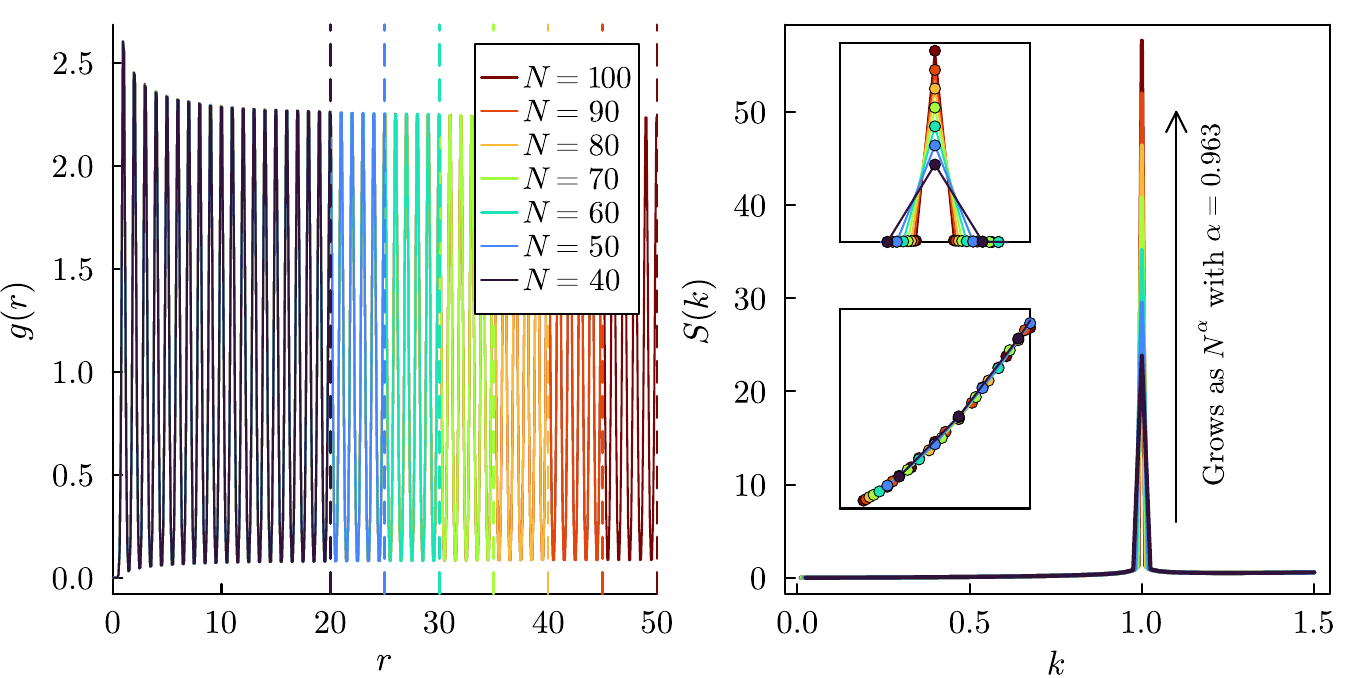}}
		\caption{We set $s = -0.5$ and $T = 0.1$, and we vary the number of particles $N$. We then regress the successive height of the peaks $S(1)$ on $N^\alpha$. Here, we found $\alpha = 0.963$, which is therefore rather close to that of a Dirac mass. }\label{fig:deter_3}
	\end{figure*} 
	
	\begin{figure*}
		\centerline{\includegraphics[width=\textwidth]{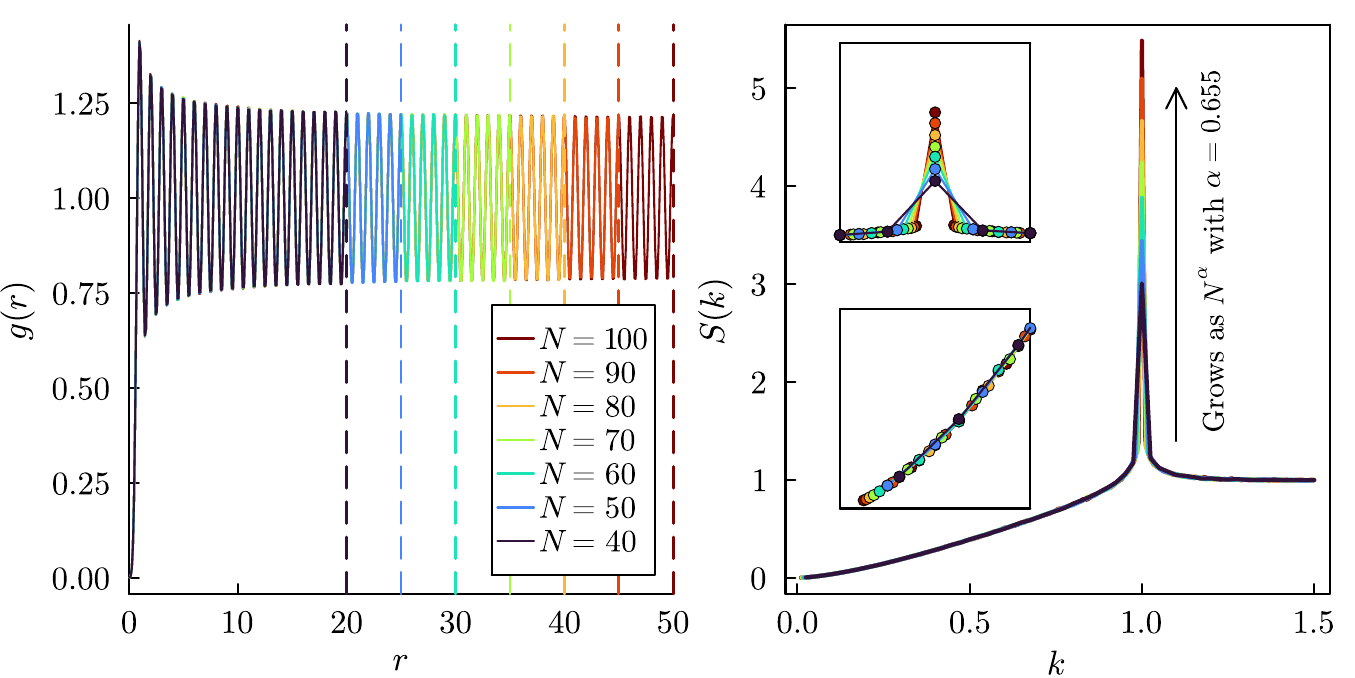}}
		\caption{Same as in \Cref{fig:deter_3}, but for $T = 0.6$. For this temperature, we have $\alpha = 0.655$, which would say that at this temperature, the Riesz gas is a quasisolid. We emphasize that it is not evident that the oscillations of the pair correlation $g(r)$ eventually vanish in the large $r$ limit, as expected for a quasisolid. This is a well-known problem in the literature, which makes the BKT-type transitions very delicate to detect numerically. }\label{fig:deter_4}
	\end{figure*} 
	
	To determine whether or not the peak at $k = 1$ is of an integrable type or a Dirac mass, we may look at its behaviour as one varies the number of particles $N$. Indeed, in the case of a Dirac mass, the height of the peak $S(1)$ should grow as $N$. On the other hand, if the structure factor diverges as $|1 - k|^{-\alpha}$ as $k \to 0$ for some $\alpha < 1$, as in a quasisolid, then the height of the peak should grow as $N^\alpha$. An example is given in \Cref{fig:deter_3} and \Cref{fig:deter_4}, in which we fix $s = -0.5$ and two different temperatures, namely $T = 0.1$ and $T = 0.6$. We then vary the number of particles and determine how the height of the peak of the structure factor grows with $N$. When $T = 0.6$, we find that the peak grows as $N^\alpha$ for $\alpha = 0.655$, which seems to indicate a quasisolid phase. When $T=0.1$, we find $\alpha = 0.963$, which is closer to indicating a Dirac mass and therefore a solid phase. 

	\subsection{Phase diagram with respect to the temperature}
	
\begin{figure*}
		\centerline{\includegraphics[width=0.7\textwidth]{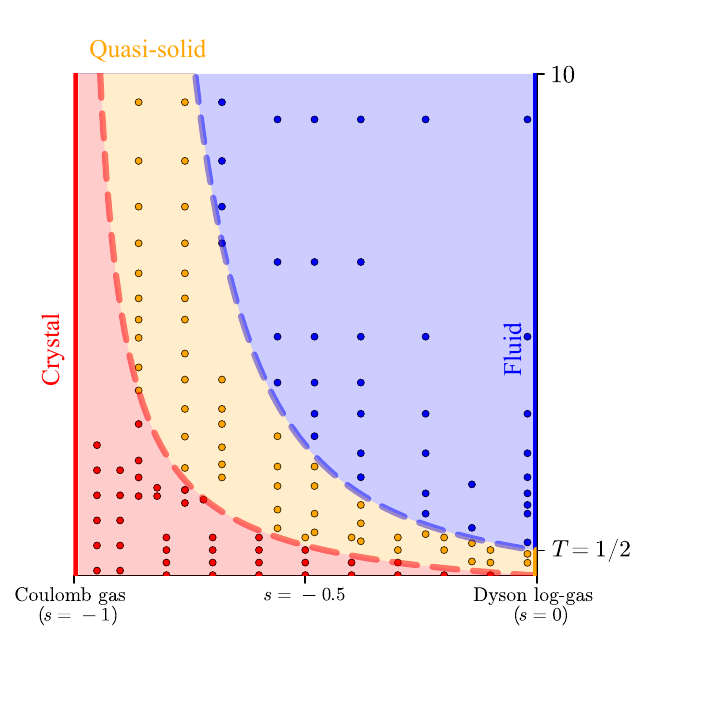}}
		\caption{Phase diagram of the one-dimensional Riesz gas with respect to the effective temperature $T$ and the exponent $-1 \leq s \leq 0$. The blue dots correspond to the couple $(s, T)$ for which we found the system to be a fluid phase. The orange dots correspond to that of the quasi-solid phase, for which the structure factor $S(k)$ has a (integrable) peak at $k = 1$ and $S(k)$ behaves linearly in the limit $k \to 0$. Finally, the red dots correspond to the solid phase, for which $S(k)$ as a Dirac mass at $k = 1$ and $S(k)$ behaves as $|k|^{1-s}$. We then draw schematically the transition curve $\widetilde{T}_s$, corresponding to the BKT transition, and the transition curve $T_s$, corresponding to the solid-phase. Those curves are seen to be consistent with the phase diagrams of the Coulomb gas and the Dyson log-gas, for which $s - 1$ and $s = 0$ respectively. }\label{fig:diagram}
	\end{figure*}

	Using the different criteria as summarized in \Cref{tab:summary}, we may draw a schematic phase diagram of the 1D Riesz gas with respect to the effective temperature $T$ and the exponent $-1 \leq s \leq 0$. The diagram is depicted in \Cref{fig:diagram}. 
	
	We should emphasize that a precise determination of the transition curves $\widetilde{T}_s$ and $T_s$ is evidently complicated. The transition curve $\widetilde{T}_s$, corresponding to what we believe to be a BKT transition separating between a fluid and a quasisolid phase, can be determined as the threshold temperature at which $S(k)$ start having a peak at $k = 1$ — thus becoming non-monotonous near $k = 1$. We see in \Cref{fig:diagram} that the behaviour of $\widetilde{T}_s$ is consistent with the Dyson log-gas for which the BKT transition occurs at $T = 1/2$.

	As for the transition curve $T_s$, corresponding to the fluid-solid transition, its values are somewhat harder to determine. According to our criterion, it corresponds to the temperature at which $S(k)$ flattens at the origin and behaves as $|k|^{1-s}$ and at which $S(k)$ as a Dirac mass at $k = 1$. Although a precise determination of this threshold is evidently complicated in the finite length $N<\infty$, we are confident that the phase diagram depicted in \Cref{fig:diagram} is qualitatively sound.

	\begin{table}
		\centering
		\caption{Summary of criteria}
		\begin{tabularx}{\linewidth}{p{0.1\linewidth}XXX}
			\toprule
			& $S(k)$ near $k = 0$         &  $S(k)$ near $k = 1$          & $g(r)$ far away           \\
			\midrule
			Solid                & \textbf{If} $S(k) \sim C|k|^\eta$ for $\eta > 1$ and some constant $C > 0$, then translational-symmety is broken according to Aizenman \textit{et al.} \cite{aizenman_bounded_2001}. Conversely, \textbf{if} the symmetry is broken then according to a heuristic of Forrester \cite{forrester_log-gases_2010}, see also \cite{alastuey_decay_1985}, it must be that $\eta = 1-s$.   & $S(k)$ has a Dirac at $k = 1$, accounting for the periodicity of $g(r)$. Numerically, the peak should grow as $N$ or equivalently its width should decrease as $N^{-1}$ & $g(r) \not\to \rho = 1$ as $r \to \infty$, and periodic in the large $r$ limit. This is reminiscent of the Coulomb gas with periodic boundary conditions \cite{kunz_one-dimensional_1974}.      \\ \midrule
			Quasi-solid              & $S(k)$ should be linear in the limit $k \to 0$. This is reminiscent of the Dyson log-gas, see \eqref{eq:S_bkt_0}.          & $S(k)$ should be non-monotonic near $k = 1$. It can be singular but must remain integrable, that is $S(k) \sim \frac{1}{|k-1|^\alpha}$ for $\alpha < 1$. Numerically, the peak should grow as $N^{\alpha}$. This is reminiscent of the Dyson log-gas, see \eqref{eq:S_bkt_1}.    & $g(r) \to \rho = 1$ as $r\to\infty$ but has oscillations which slowly vanish in the large $r$ limit. The power of leading order term should depend on the temperature.     \\ \midrule
			Fluid   &   $S(k)$ behaves linearly in the limit $k\to0$.     & $S(k)$ is monotonic near $k=1$     & $g(r) \sim 1 - 1/r^2$ in the large $r$ limit.       \\
			\bottomrule
		\end{tabularx}
		\label{tab:summary}
	\end{table}

	\section{Conclusion}
	We provided numerical evidence for the existence of two distinct phase transitions with respect to the temperature in 1D Riesz gases. The first transition corresponds to a BKT transition similar to that of the Dyson log-gas. The second one corresponds to the critical temperature below which the system is crystallized. The transition curves interpolate between the known — or at least expected — phase diagrams of the 1D Coulomb gas and the Dyson log-gas. Although a precise quantitative determination on the critical temperatures is delicate, using a set of different criteria we were able to draw a schematic phase diagram of the 1D Riesz gases with respect to the temperature $T$ and the exponent $s$. 
	\bmhead{Acknowledgments}
	The author is thankful to Mathieu Lewin (\textsc{CNRS \& Ceremade}, Université Paris--Dauphine PSL) as well as David Dereudre (Université de Lille) for useful discussions. This project has partially received funding from the 
	European Research Council (ERC) under the European Union's Horizon 2020 research and innovation program (grant agreement MDFT N°725528).
	
	\appendix
	\section{Implementation}\label{app}
	We very briefly comment on the algorithm used in this paper. Our code was written in \texttt{Julia}\footnote{Our code is available at \url{https://github.com/rodriguel/PTRiesz}.}. The 1D Riesz gases were simulated using a \emph{random walk Metropolis-Hastings} with appropriate tuning of the size of the jump proposal to achieve a good acceptance rate. A cluster architecture was used to produce many samples in a parallel fashion. The periodic Riesz potential $\widetilde{v}_{s, L}$ was pre-computed by tabulation and interpolation so as not to use special functions whose evaluation are rather time consuming. The pair correlation $g(r)$ is obtained by binning. That is, for each sample generated by the chain, the mutual distances between the particles in the configurations are computed, and those distances are binned into an histogram which is then properly normalized. The number of bins should evidently depend of the number of samples used and the accuracy needed. As a general rule, we chose to consider ten bins \emph{per} unit length. We actually found this to be usually consistent with the \emph{Freedman--Diaconis rule} \cite{freedman_histogram_1981}. To the best of our knowledge, binning the pair correlation seems to be the most commonly used method, at the exception of the work \cite{patrone_beyond_2017}. Once the pair correlation has been properly binned into an histogram, we compute the structure factor $S(k)$ by considering the (discrete) Fourier transform of this histogram, which is then only defined on the values $\Z/N$. Once again, this seems to be the usual procedure in the literature to compute $S(k)$ \cite{patrone_beyond_2017, hawat_estimating_2023}. Computations were carried out using a number of particles up to $N \sim 500$, and we found no significant differences beyond intrinsic noise with the less greedy choice of $N \sim 100$, so that most computations in this paper were carried out for a number of particles $N$ of this order. 
	


\begin{thebibliography}{72}
\ifx \bisbn   \undefined \def \bisbn  #1{ISBN #1}\fi
\ifx \binits  \undefined \def \binits#1{#1}\fi
\ifx \bauthor  \undefined \def \bauthor#1{#1}\fi
\ifx \batitle  \undefined \def \batitle#1{#1}\fi
\ifx \bjtitle  \undefined \def \bjtitle#1{#1}\fi
\ifx \bvolume  \undefined \def \bvolume#1{\textbf{#1}}\fi
\ifx \byear  \undefined \def \byear#1{#1}\fi
\ifx \bissue  \undefined \def \bissue#1{#1}\fi
\ifx \bfpage  \undefined \def \bfpage#1{#1}\fi
\ifx \blpage  \undefined \def \blpage #1{#1}\fi
\ifx \burl  \undefined \def \burl#1{\textsf{#1}}\fi
\ifx \doiurl  \undefined \def \doiurl#1{\url{https://doi.org/#1}}\fi
\ifx \betal  \undefined \def \betal{\textit{et al.}}\fi
\ifx \binstitute  \undefined \def \binstitute#1{#1}\fi
\ifx \binstitutionaled  \undefined \def \binstitutionaled#1{#1}\fi
\ifx \bctitle  \undefined \def \bctitle#1{#1}\fi
\ifx \beditor  \undefined \def \beditor#1{#1}\fi
\ifx \bpublisher  \undefined \def \bpublisher#1{#1}\fi
\ifx \bbtitle  \undefined \def \bbtitle#1{#1}\fi
\ifx \bedition  \undefined \def \bedition#1{#1}\fi
\ifx \bseriesno  \undefined \def \bseriesno#1{#1}\fi
\ifx \blocation  \undefined \def \blocation#1{#1}\fi
\ifx \bsertitle  \undefined \def \bsertitle#1{#1}\fi
\ifx \bsnm \undefined \def \bsnm#1{#1}\fi
\ifx \bsuffix \undefined \def \bsuffix#1{#1}\fi
\ifx \bparticle \undefined \def \bparticle#1{#1}\fi
\ifx \barticle \undefined \def \barticle#1{#1}\fi
\bibcommenthead
\ifx \bconfdate \undefined \def \bconfdate #1{#1}\fi
\ifx \botherref \undefined \def \botherref #1{#1}\fi
\ifx \url \undefined \def \url#1{\textsf{#1}}\fi
\ifx \bchapter \undefined \def \bchapter#1{#1}\fi
\ifx \bbook \undefined \def \bbook#1{#1}\fi
\ifx \bcomment \undefined \def \bcomment#1{#1}\fi
\ifx \oauthor \undefined \def \oauthor#1{#1}\fi
\ifx \citeauthoryear \undefined \def \citeauthoryear#1{#1}\fi
\ifx \endbibitem  \undefined \def \endbibitem {}\fi
\ifx \bconflocation  \undefined \def \bconflocation#1{#1}\fi
\ifx \arxivurl  \undefined \def \arxivurl#1{\textsf{#1}}\fi
\csname PreBibitemsHook\endcsname

\bibitem[\protect\citeauthoryear{Hohenberg}{}]{hohenberg_existence_1967}
\begin{botherref}
\oauthor{\bsnm{Hohenberg}, \binits{P.C.}}:
Existence of long-range order in one and two dimensions
\textbf{158}(2),
383--386
\doiurl{10.1103/physrev.158.383}
\end{botherref}
\endbibitem

\bibitem[\protect\citeauthoryear{Mermin and Wagner}{}]{mermin_absence_1966}
\begin{botherref}
\oauthor{\bsnm{Mermin}, \binits{N.D.}},
\oauthor{\bsnm{Wagner}, \binits{H.}}:
Absence of ferromagnetism or antiferromagnetism in one- or two-dimensional
  isotropic heisenberg models
\textbf{17}(22),
1133--1136
\doiurl{10.1103/physrevlett.17.1133}
\end{botherref}
\endbibitem

\bibitem[\protect\citeauthoryear{Mermin}{}]{mermin_absence_1967}
\begin{botherref}
\oauthor{\bsnm{Mermin}, \binits{N.D.}}:
Absence of ordering in certain classical systems
\textbf{8}(5),
1061--1064
\doiurl{10.1063/1.1705316}
\end{botherref}
\endbibitem

\bibitem[\protect\citeauthoryear{Mermin}{}]{mermin_crystalline_1968}
\begin{botherref}
\oauthor{\bsnm{Mermin}, \binits{N.D.}}:
Crystalline order in two dimensions
\textbf{176}(1),
250--254
\doiurl{10.1103/PhysRev.176.250}
\end{botherref}
\endbibitem

\bibitem[\protect\citeauthoryear{Van~Hove}{}]{van_hove_sur_1950}
\begin{botherref}
\oauthor{\bsnm{Van~Hove}, \binits{L.}}:
Sur l'intégrale de configuration pour les systèmes de particules à une
  dimension
\textbf{16}(2),
137--143
\end{botherref}
\endbibitem

\bibitem[\protect\citeauthoryear{Dyson}{}]{dyson_existence_1969}
\begin{botherref}
\oauthor{\bsnm{Dyson}, \binits{F.J.}}:
Existence of a phase-transition in a one-dimensional {{Ising}} ferromagnet
\textbf{12}(2),
91--107
\doiurl{10.1007/BF01645907}
\end{botherref}
\endbibitem

\bibitem[\protect\citeauthoryear{Kittel}{}]{kittel_phase_1969}
\begin{botherref}
\oauthor{\bsnm{Kittel}, \binits{C.}}:
Phase transition of a molecular zipper
\textbf{37}(9),
917--920
\doiurl{10.1119/1.1975930}
\end{botherref}
\endbibitem

\bibitem[\protect\citeauthoryear{Chui and Weeks}{}]{chui_pinning_1981}
\begin{botherref}
\oauthor{\bsnm{Chui}, \binits{S.T.}},
\oauthor{\bsnm{Weeks}, \binits{J.D.}}:
Pinning and roughening of one-dimensional models of interfaces and steps
\textbf{23}(5),
2438--2441
\doiurl{10.1103/PhysRevB.23.2438}
\end{botherref}
\endbibitem

\bibitem[\protect\citeauthoryear{Großkinsky
  et~al.}{}]{groskinsky_condensation_2003}
\begin{botherref}
\oauthor{\bsnm{Großkinsky}, \binits{S.}},
\oauthor{\bsnm{Schütz}, \binits{G.M.}},
\oauthor{\bsnm{Spohn}, \binits{H.}}:
Condensation in the zero range process: {{Stationary}} and dynamical properties
\textbf{113}(3),
389--410
\end{botherref}
\endbibitem

\bibitem[\protect\citeauthoryear{Sarkanych et~al.}{}]{sarkanych_exact_2017}
\begin{botherref}
\oauthor{\bsnm{Sarkanych}, \binits{P.}},
\oauthor{\bsnm{Holovatch}, \binits{Y.}},
\oauthor{\bsnm{Kenna}, \binits{R.}}:
Exact solution of a classical short-range spin model with a phase transition in
  one dimension: {{The Potts}} model with invisible states
\textbf{381}(41),
3589--3593
\doiurl{10.1016/j.physleta.2017.08.063}
\end{botherref}
\endbibitem

\bibitem[\protect\citeauthoryear{Saryal et~al.}{}]{saryal_multiple_2018}
\begin{botherref}
\oauthor{\bsnm{Saryal}, \binits{S.}},
\oauthor{\bsnm{Klamser}, \binits{J.U.}},
\oauthor{\bsnm{Sadhu}, \binits{T.}},
\oauthor{\bsnm{Dhar}, \binits{D.}}:
Multiple {{Singularities}} of the {{Equilibrium Free Energy}} in a
  {{One-Dimensional Model}} of {{Soft Rods}}
\textbf{121}(24),
240601
\doiurl{10.1103/PhysRevLett.121.240601}
\end{botherref}
\endbibitem

\bibitem[\protect\citeauthoryear{Cuesta and Sánchez}{}]{cuesta_general_2004}
\begin{botherref}
\oauthor{\bsnm{Cuesta}, \binits{J.A.}},
\oauthor{\bsnm{Sánchez}, \binits{A.}}:
General non-existence theorem for phase transitions in one-dimensional systems
  with short range interactions, and physical examples of such transitions
\textbf{115}(3-4),
869--893
\doiurl{10.1023/B:JOSS.0000022373.63640.4e}
\end{botherref}
\endbibitem

\bibitem[\protect\citeauthoryear{Kunz}{}]{kunz_one-dimensional_1974}
\begin{botherref}
\oauthor{\bsnm{Kunz}, \binits{H.}}:
The one-dimensional classical electron gas
\textbf{85}(2),
303--335
\doiurl{10.1016/0003-4916(74)90413-8}
\end{botherref}
\endbibitem

\bibitem[\protect\citeauthoryear{Aizenman and
  Martin}{}]{aizenman_structure_1980}
\begin{botherref}
\oauthor{\bsnm{Aizenman}, \binits{M.}},
\oauthor{\bsnm{Martin}, \binits{P.A.}}:
Structure of gibbs states of one dimensional {{Coulomb}} systems
\textbf{78}(1),
99--116
\doiurl{10.1007/BF01941972}
\end{botherref}
\endbibitem

\bibitem[\protect\citeauthoryear{Brascamp and
  Lieb}{}]{brascamp_inequalities_2002}
\begin{botherref}
\oauthor{\bsnm{Brascamp}, \binits{H.J.}},
\oauthor{\bsnm{Lieb}, \binits{E.H.}}:
Some inequalities for {{Gaussian}} measures and the long-range order of the
  one-dimensional plasma.
In: Inequalities: {{Selecta}} of {{Elliott H}}. {{Lieb}},
pp. 403--416.
{Springer}.
\doiurl{10.1007/978-3-642-55925-9_34}
\end{botherref}
\endbibitem

\bibitem[\protect\citeauthoryear{Jansen and Jung}{}]{jansen_wigner_2014}
\begin{botherref}
\oauthor{\bsnm{Jansen}, \binits{S.}},
\oauthor{\bsnm{Jung}, \binits{P.}}:
Wigner crystallization in the quantum {{1D}} jellium at all densities
\textbf{331}(3),
1133--1154
\doiurl{10.1007/s00220-014-2032-y}
\end{botherref}
\endbibitem

\bibitem[\protect\citeauthoryear{Johansson}{}]{johansson_separation_1991}
\begin{botherref}
\oauthor{\bsnm{Johansson}, \binits{K.}}:
Separation of phases at low temperatures in a one-dimensional continuous gas
\textbf{141}(2),
259--278
\doiurl{10.1007/BF02101505}
\end{botherref}
\endbibitem

\bibitem[\protect\citeauthoryear{Johansson}{}]{johansson_separation_1995}
\begin{botherref}
\oauthor{\bsnm{Johansson}, \binits{K.}}:
On separation of phases in one-dimensional gases
\textbf{169}(3),
521--561
\doiurl{10.1007/BF02099311}
\end{botherref}
\endbibitem

\bibitem[\protect\citeauthoryear{Lewin}{}]{lewin_coulomb_2022}
\begin{botherref}
\oauthor{\bsnm{Lewin}, \binits{M.}}:
Coulomb and {{Riesz}} gases: {{The}} known and the unknown
\textbf{63}(6),
061101
\doiurl{10.1063/5.0086835}
\end{botherref}
\endbibitem

\bibitem[\protect\citeauthoryear{Kosterlitz and
  Thouless}{}]{kosterlitz_ordering_1973}
\begin{botherref}
\oauthor{\bsnm{Kosterlitz}, \binits{J.M.}},
\oauthor{\bsnm{Thouless}, \binits{D.J.}}:
Ordering, metastability and phase transitions in two-dimensional systems
\textbf{6}(7),
1181
\doiurl{10.1088/0022-3719/6/7/010}
\end{botherref}
\endbibitem

\bibitem[\protect\citeauthoryear{Herbut}{}]{herbut_modern_2007}
\begin{botherref}
\oauthor{\bsnm{Herbut}, \binits{I.}}:
A Modern Approach to Critical Phenomena.
\doiurl{10.2277/0521854520}
\end{botherref}
\endbibitem

\bibitem[\protect\citeauthoryear{Mudry}{}]{mudry_lecture_2014}
\begin{botherref}
\oauthor{\bsnm{Mudry}, \binits{C.}}:
Lecture Notes on Field Theory in Condensed Matter Physics.
\doiurl{10.1142/8697}
\end{botherref}
\endbibitem

\bibitem[\protect\citeauthoryear{Bernasconi and
  Schneider}{}]{bernasconi_physics_2012}
\begin{botherref}
\oauthor{\bsnm{Bernasconi}, \binits{J.}},
\oauthor{\bsnm{Schneider}, \binits{T.}}:
Physics in One Dimension: {{Proceedings}} of an International Conference
  Fribourg, Switzerland, August 25–29, 1980
vol. 23.
{Springer Science \& Business Media}
\end{botherref}
\endbibitem

\bibitem[\protect\citeauthoryear{Lieb and Mattis}{}]{lieb_mathetical_1967}
\begin{botherref}
\oauthor{\bsnm{Lieb}, \binits{E.H.}},
\oauthor{\bsnm{Mattis}, \binits{D.C.}}:
Mathematical physics in one dimension
\textbf{35}(9),
895--896
\doiurl{10.1119/1.1974285}
\end{botherref}
\endbibitem

\bibitem[\protect\citeauthoryear{Aizenman et~al.}{}]{aizenman_symmetry_2010}
\begin{botherref}
\oauthor{\bsnm{Aizenman}, \binits{M.}},
\oauthor{\bsnm{Jansen}, \binits{S.}},
\oauthor{\bsnm{Jung}, \binits{P.}}:
Symmetry breaking in quasi-{{1D Coulomb}} systems
\textbf{11}(8),
1453--1485
\doiurl{10.1007/s00023-010-0067-y}
\end{botherref}
\endbibitem

\bibitem[\protect\citeauthoryear{Chafaï et~al.}{}]{chafai_at_2020}
\begin{botherref}
\oauthor{\bsnm{Chafaï}, \binits{D.}},
\oauthor{\bsnm{García-Zelada}, \binits{D.}},
\oauthor{\bsnm{Jung}, \binits{P.}}:
At the edge of a one-dimensional jellium,
2012--04633
{\href{https://arxiv.org/abs/2012.04633}{{2012.04633}}}
\end{botherref}
\endbibitem

\bibitem[\protect\citeauthoryear{Choquard}{}]{choquard_statistical_1975}
\begin{botherref}
\oauthor{\bsnm{Choquard}, \binits{P.}}:
On the statistical mechanics of one-dimensional {{Coulomb}} systems
\textbf{48}(4),
585--598
\end{botherref}
\endbibitem

\bibitem[\protect\citeauthoryear{Dyson}{}]{dyson_statistical_1962}
\begin{botherref}
\oauthor{\bsnm{Dyson}, \binits{F.J.}}:
Statistical theory of the energy levels of complex systems. {{I}}
\textbf{3}(1),
140--156
\doiurl{10.1063/1.1703773}
\end{botherref}
\endbibitem

\bibitem[\protect\citeauthoryear{Dyson}{}]{dyson_statistical_1962-1}
\begin{botherref}
\oauthor{\bsnm{Dyson}, \binits{F.J.}}:
Statistical theory of the energy levels of complex systems. {{II}}
\textbf{3}(1),
157--165
\doiurl{10.1063/1.1703774}
\end{botherref}
\endbibitem

\bibitem[\protect\citeauthoryear{Dyson}{}]{dyson_statistical_1962-2}
\begin{botherref}
\oauthor{\bsnm{Dyson}, \binits{F.J.}}:
Statistical theory of the energy levels of complex systems. {{III}}
\textbf{3}(1),
166--175
\doiurl{10.1063/1.1703775}
\end{botherref}
\endbibitem

\bibitem[\protect\citeauthoryear{Dyson and Mehta}{}]{dyson_statistical_1963}
\begin{botherref}
\oauthor{\bsnm{Dyson}, \binits{F.J.}},
\oauthor{\bsnm{Mehta}, \binits{M.L.}}:
Statistical theory of the energy levels of complex systems. {{IV}}
\textbf{4}(5),
701--712
\doiurl{10.1063/1.1704008}
\end{botherref}
\endbibitem

\bibitem[\protect\citeauthoryear{Forrester}{}]{forrester_log-gases_2010}
\begin{botherref}
\oauthor{\bsnm{Forrester}, \binits{P.J.}}:
Log-{{Gases}} and {{Random Matrices}} ({{LMS-34}}).
In: Log-{{Gases}} and {{Random Matrices}} ({{LMS-34}}).
{Princeton University Press}.
\doiurl{10.1515/9781400835416}
\end{botherref}
\endbibitem

\bibitem[\protect\citeauthoryear{Valkó and Virág}{}]{valko_continuum_2009-4}
\begin{botherref}
\oauthor{\bsnm{Valkó}, \binits{B.}},
\oauthor{\bsnm{Virág}, \binits{B.}}:
Continuum limits of random matrices and~the~{{Brownian}}~carousel
\textbf{177}(3),
463--508
\doiurl{10.1007/s00222-009-0180-z}
\end{botherref}
\endbibitem

\bibitem[\protect\citeauthoryear{Sandier and Serfaty}{}]{sandier_1d_2015}
\begin{botherref}
\oauthor{\bsnm{Sandier}, \binits{E.}},
\oauthor{\bsnm{Serfaty}, \binits{S.}}:
{{1D}} log gases and the renormalized energy: {{Crystallization}} at vanishing
  temperature
\textbf{162}(3),
795--846
\doiurl{10.1007/s00440-014-0585-5}
\end{botherref}
\endbibitem

\bibitem[\protect\citeauthoryear{Requardt and Wagner}{}]{requardt_wigner_1990}
\begin{botherref}
\oauthor{\bsnm{Requardt}, \binits{M.}},
\oauthor{\bsnm{Wagner}, \binits{H.J.}}:
Wigner crystallization and its relation to the poor decay of pair correlations
  in one-component plasmas of arbitrary dimension
\textbf{58}(5),
1165--1180
\doiurl{10.1007/BF01026570}
\end{botherref}
\endbibitem

\bibitem[\protect\citeauthoryear{Erbar et~al.}{}]{erbar_one-dimensional_2021}
\begin{botherref}
\oauthor{\bsnm{Erbar}, \binits{M.}},
\oauthor{\bsnm{Huesmann}, \binits{M.}},
\oauthor{\bsnm{Leblé}, \binits{T.}}:
The {{One-Dimensional Log-Gas Free Energy Has}} a {{Unique Minimizer}}
\textbf{74}(3),
615--675
\doiurl{10.1002/cpa.21977}
\end{botherref}
\endbibitem

\bibitem[\protect\citeauthoryear{Dereudre}{}]{dereudre_introduction_2019}
\begin{botherref}
\oauthor{\bsnm{Dereudre}, \binits{D.}}:
Introduction to the theory of {{Gibbs}} point processes.
In: Stochastic Geometry.
Lecture Notes in Math.,
vol. 2237,
pp. 181--229.
{Springer, Cham}
\end{botherref}
\endbibitem

\bibitem[\protect\citeauthoryear{Papangelou}{}]{papangelou_absence_1987}
\begin{botherref}
\oauthor{\bsnm{Papangelou}, \binits{F.}}:
On the absence of phase transition in continuous one-dimensional {{Gibbs}}
  systems with no hard core
\textbf{74}(4),
485--496
\doiurl{10.1007/BF00363511}
\end{botherref}
\endbibitem

\bibitem[\protect\citeauthoryear{Hardin et~al.}{}]{hardin_next_2015}
\begin{botherref}
\oauthor{\bsnm{Hardin}, \binits{D.P.}},
\oauthor{\bsnm{Saff}, \binits{E.B.}},
\oauthor{\bsnm{Simanek}, \binits{B.Z.}},
\oauthor{\bsnm{Su}, \binits{Y.}}:
Next order energy asymptotics for {{Riesz}} potentials on flat tori,
1511--01552
{\href{https://arxiv.org/abs/1511.01552}{{1511.01552}}}
\end{botherref}
\endbibitem

\bibitem[\protect\citeauthoryear{Borodachov
  et~al.}{}]{borodachov_discrete_2019}
\begin{botherref}
\oauthor{\bsnm{Borodachov}, \binits{S.V.}},
\oauthor{\bsnm{Hardin}, \binits{D.P.}},
\oauthor{\bsnm{Saff}, \binits{E.B.}}:
Discrete Energy on Rectifiable Sets.
{Springer}
\end{botherref}
\endbibitem

\bibitem[\protect\citeauthoryear{Serfaty}{}]{serfaty_coulomb_2015}
\begin{botherref}
\oauthor{\bsnm{Serfaty}, \binits{S.}}:
Coulomb {{Gases}} and {{Ginzburg}}–{{Landau Vortices}}.
{EMS}.
\doiurl{10.4171/152}
\end{botherref}
\endbibitem

\bibitem[\protect\citeauthoryear{Ruelle}{}]{ruelle_statistical_1999}
\begin{botherref}
\oauthor{\bsnm{Ruelle}, \binits{D.}}:
Statistical {{Mechanics}}: {{Rigorous Results}}.
{World Scientific}
\end{botherref}
\endbibitem

\bibitem[\protect\citeauthoryear{Georgii}{}]{georgii_canonical_1976}
\begin{botherref}
\oauthor{\bsnm{Georgii}, \binits{H.-O.}}:
Canonical and grand canonical {{Gibbs}} states for continuum systems
\textbf{48}(1),
31--51
\doiurl{10.1007/BF01609410}
\end{botherref}
\endbibitem

\bibitem[\protect\citeauthoryear{Georgii}{}]{georgii_gibbs_2011}
\begin{botherref}
\oauthor{\bsnm{Georgii}, \binits{H.-O.}}:
Gibbs {{Measures}} and {{Phase Transitions}}.
In: Gibbs {{Measures}} and {{Phase Transitions}}.
{De Gruyter}.
\doiurl{10.1515/9783110250329}
\end{botherref}
\endbibitem

\bibitem[\protect\citeauthoryear{Doerushin and
  Minlos}{}]{doerushin_existence_1967}
\begin{botherref}
\oauthor{\bsnm{Doerushin}, \binits{R.L.}},
\oauthor{\bsnm{Minlos}, \binits{R.A.}}:
Existence and {{Continuity}} of {{Pressure}} in {{Classical Statistical
  Physics}}
\textbf{12}(4),
535--559
\doiurl{10.1137/1112072}
\end{botherref}
\endbibitem

\bibitem[\protect\citeauthoryear{Dereudre and
  Vasseur}{}]{dereudre_number-rigidity_2023}
\begin{botherref}
\oauthor{\bsnm{Dereudre}, \binits{D.}},
\oauthor{\bsnm{Vasseur}, \binits{T.}}:
Number-rigidity and beta-circular {{Riesz}} gas
\textbf{51}(3),
1025--1065
\doiurl{10.1214/22-AOP1606}
\end{botherref}
\endbibitem

\bibitem[\protect\citeauthoryear{Boursier}{}]{boursier_decay_2022}
\begin{botherref}
\oauthor{\bsnm{Boursier}, \binits{J.}}:
Decay of Correlations and Thermodynamic Limit for the Circular {{Riesz}} Gas
\end{botherref}
\endbibitem

\bibitem[\protect\citeauthoryear{Dereudre et~al.}{}]{dereudre_dlr_2021-1}
\begin{botherref}
\oauthor{\bsnm{Dereudre}, \binits{D.}},
\oauthor{\bsnm{Hardy}, \binits{A.}},
\oauthor{\bsnm{Leblé}, \binits{T.}},
\oauthor{\bsnm{Maïda}, \binits{M.}}:
{{DLR Equations}} and {{Rigidity}} for the {{Sine-Beta Process}}
\textbf{74}(1),
172--222
\doiurl{10.1002/cpa.21963}
\end{botherref}
\endbibitem

\bibitem[\protect\citeauthoryear{Lenard}{}]{lenard_exact_2004}
\begin{botherref}
\oauthor{\bsnm{Lenard}, \binits{A.}}:
Exact {{Statistical Mechanics}} of a {{One}}‐{{Dimensional System}} with
  {{Coulomb Forces}}. {{III}}. {{Statistics}} of the {{Electric Field}}
\textbf{4}(4),
533--543
\doiurl{10.1063/1.1703988}
\end{botherref}
\endbibitem

\bibitem[\protect\citeauthoryear{Hansen and McDonald}{}]{hansen_theory_2013}
\begin{botherref}
\oauthor{\bsnm{Hansen}, \binits{J.-P.}},
\oauthor{\bsnm{McDonald}, \binits{I.R.}}:
Theory of {{Simple Liquids}}.
{Academic Press}.
\doiurl{10.1016/B978-0-12-387032-2.00013-1}
\end{botherref}
\endbibitem

\bibitem[\protect\citeauthoryear{Dinnebier and
  Billinge}{}]{dinnebier_powder_2008}
\begin{botherref}
\oauthor{\bsnm{Dinnebier}, \binits{R.}},
\oauthor{\bsnm{Billinge}, \binits{S.}}:
Powder {{Diffraction}}: {{Theory}} And {{Practice}}.
\doiurl{10.1039/9781847558237}
\end{botherref}
\endbibitem

\bibitem[\protect\citeauthoryear{Gingrich and
  Heaton}{}]{gingrich_structure_2004}
\begin{botherref}
\oauthor{\bsnm{Gingrich}, \binits{N.S.}},
\oauthor{\bsnm{Heaton}, \binits{L.}}:
Structure of {{Alkali Metals}} in the {{Liquid State}}
\textbf{34}(3),
873--878
\doiurl{10.1063/1.1731688}
\end{botherref}
\endbibitem

\bibitem[\protect\citeauthoryear{Sirota et~al.}{}]{sirota_complete_1989}
\begin{botherref}
\oauthor{\bsnm{Sirota}, \binits{E.B.}},
\oauthor{\bsnm{Ou-Yang}, \binits{H.D.}},
\oauthor{\bsnm{Sinha}, \binits{S.K.}},
\oauthor{\bsnm{Chaikin}, \binits{P.M.}},
\oauthor{\bsnm{Axe}, \binits{J.D.}},
\oauthor{\bsnm{Fujii}, \binits{Y.}}:
Complete phase diagram of a charged colloidal system: {{A}} synchro- tron
  x-{{Ray}} scattering study
\textbf{62}(13),
1524--1527
\doiurl{10.1103/PhysRevLett.62.1524}
\end{botherref}
\endbibitem

\bibitem[\protect\citeauthoryear{Yarnell et~al.}{}]{yarnell_structure_1973}
\begin{botherref}
\oauthor{\bsnm{Yarnell}, \binits{J.L.}},
\oauthor{\bsnm{Katz}, \binits{M.J.}},
\oauthor{\bsnm{Wenzel}, \binits{R.G.}},
\oauthor{\bsnm{Koenig}, \binits{S.H.}}:
Structure {{Factor}} and {{Radial Distribution Function}} for {{Liquid Argon}}
  at {{85K}}
\textbf{7}(6),
2130--2144
\doiurl{10.1103/PhysRevA.7.2130}
\end{botherref}
\endbibitem

\bibitem[\protect\citeauthoryear{Borwein et~al.}{}]{borwein_convergence_1985}
\begin{botherref}
\oauthor{\bsnm{Borwein}, \binits{D.}},
\oauthor{\bsnm{Borwein}, \binits{J.M.}},
\oauthor{\bsnm{Taylor}, \binits{K.F.}}:
Convergence of lattice sums and {{Madelung}}’s constant
\textbf{26}(11),
2999--3009
\doiurl{10.1063/1.526675}
\end{botherref}
\endbibitem

\bibitem[\protect\citeauthoryear{Borwein et~al.}{}]{borwein_energy_1988}
\begin{botherref}
\oauthor{\bsnm{Borwein}, \binits{D.}},
\oauthor{\bsnm{Borwein}, \binits{J.M.}},
\oauthor{\bsnm{Shail}, \binits{R.}},
\oauthor{\bsnm{Zucker}, \binits{I.J.}}:
Energy of static electron lattices
\textbf{21}(7),
1519
\doiurl{10.1088/0305-4470/21/7/015}
\end{botherref}
\endbibitem

\bibitem[\protect\citeauthoryear{Borwein et~al.}{}]{borwein_analysis_1989}
\begin{botherref}
\oauthor{\bsnm{Borwein}, \binits{D.}},
\oauthor{\bsnm{Borwein}, \binits{J.M.}},
\oauthor{\bsnm{Shail}, \binits{R.}}:
Analysis of certain lattice sums
\textbf{143}(1),
126--137
\doiurl{10.1016/0022-247X(89)90032-2}
\end{botherref}
\endbibitem

\bibitem[\protect\citeauthoryear{Borwein et~al.}{}]{borwein_lattice_2014}
\begin{botherref}
\oauthor{\bsnm{Borwein}, \binits{D.}},
\oauthor{\bsnm{Borwein}, \binits{J.M.}},
\oauthor{\bsnm{Straub}, \binits{A.}}:
On lattice sums and {{Wigner}} limits
\textbf{414}(2),
489--513
\doiurl{10.1016/j.jmaa.2014.01.008}
\end{botherref}
\endbibitem

\bibitem[\protect\citeauthoryear{Baxter}{}]{baxter_exactly_2016}
\begin{botherref}
\oauthor{\bsnm{Baxter}, \binits{R.J.}}:
Exactly {{Solved Models}} in {{Statistical Mechanics}}.
{Elsevier}
\end{botherref}
\endbibitem

\bibitem[\protect\citeauthoryear{Lenard}{}]{lenard_exact_2004-1}
\begin{botherref}
\oauthor{\bsnm{Lenard}, \binits{A.}}:
Exact {{Statistical Mechanics}} of a {{One}}‐{{Dimensional System}} with
  {{Coulomb Forces}}
\textbf{2}(5),
682--693
\doiurl{10.1063/1.1703757}
\end{botherref}
\endbibitem

\bibitem[\protect\citeauthoryear{Edwards and Lenard}{}]{edwards_exact_2004}
\begin{botherref}
\oauthor{\bsnm{Edwards}, \binits{S.F.}},
\oauthor{\bsnm{Lenard}, \binits{A.}}:
Exact {{Statistical Mechanics}} of a {{One}}‐{{Dimensional System}} with
  {{Coulomb Forces}}. {{II}}. {{The Method}} of {{Functional Integration}}
\textbf{3}(4),
778--792
\doiurl{10.1063/1.1724281}
\end{botherref}
\endbibitem

\bibitem[\protect\citeauthoryear{Haldane}{}]{haldane_effective_1981}
\begin{botherref}
\oauthor{\bsnm{Haldane}, \binits{F.D.M.}}:
Effective harmonic-fluid approach to low-energy properties of one-dimensional
  quantum fluids
\textbf{47}(25),
1840--1843
\doiurl{10.1103/PhysRevLett.47.1840}
\end{botherref}
\endbibitem

\bibitem[\protect\citeauthoryear{Forrester}{}]{forrester_analogues_1984}
\begin{botherref}
\oauthor{\bsnm{Forrester}, \binits{P.J.}}:
Analogues between a quantum many body problem and the log-gas
\textbf{17}(10),
2059--2067
\end{botherref}
\endbibitem

\bibitem[\protect\citeauthoryear{Forrester}{}]{forrester_exact_1993}
\begin{botherref}
\oauthor{\bsnm{Forrester}, \binits{P.J.}}:
Exact integral formulas and asymptotics for the correlations in the {$1/r^2$}
  quantum many body system
\textbf{179}(2),
127--130
\doiurl{10.1016/0375-9601(93)90661-I}
\end{botherref}
\endbibitem

\bibitem[\protect\citeauthoryear{Leblé}{}]{leble_uniqueness_2015}
\begin{botherref}
\oauthor{\bsnm{Leblé}, \binits{T.}}:
A uniqueness result for minimizers of the {{1D Log-gas}} renormalized energy
\textbf{268}(7),
1649--1677
\doiurl{10.1016/j.jfa.2014.11.023}
\end{botherref}
\endbibitem

\bibitem[\protect\citeauthoryear{Aizenman et~al.}{}]{aizenman_bounded_2001}
\begin{botherref}
\oauthor{\bsnm{Aizenman}, \binits{M.}},
\oauthor{\bsnm{Goldstein}, \binits{S.}},
\oauthor{\bsnm{Lebowitz}, \binits{J.L.}}:
Bounded fluctuations and translation symmetry breaking in one-dimensional
  particle systems.
In: Journal of {{Statistical Physics}}
vol. 103,
pp. 601--618.
\doiurl{10.1023/A:1010397401128}
\end{botherref}
\endbibitem

\bibitem[\protect\citeauthoryear{Torquato and
  Stillinger}{}]{torquato_local_2003}
\begin{botherref}
\oauthor{\bsnm{Torquato}, \binits{S.}},
\oauthor{\bsnm{Stillinger}, \binits{F.H.}}:
Local density fluctuations, hyperuniformity, and order metrics
\textbf{68}(4),
041113
\doiurl{10.1103/PhysRevE.68.041113}
\end{botherref}
\endbibitem

\bibitem[\protect\citeauthoryear{Torquato}{}]{torquato_hyperuniform_2018}
\begin{botherref}
\oauthor{\bsnm{Torquato}, \binits{S.}}:
Hyperuniform states of matter
\textbf{745},
1--95
\doiurl{10.1016/j.physrep.2018.03.001}
\end{botherref}
\endbibitem

\bibitem[\protect\citeauthoryear{Alastuey and Martin}{}]{alastuey_decay_1985}
\begin{botherref}
\oauthor{\bsnm{Alastuey}, \binits{A.}},
\oauthor{\bsnm{Martin}, \binits{P.-A.}}:
Decay of correlations in classical fluids with long-range forces
\textbf{39}(3-4),
405--426
\doiurl{10.1007/BF01018670}
\end{botherref}
\endbibitem

\bibitem[\protect\citeauthoryear{Freedman and
  Diaconis}{}]{freedman_histogram_1981}
\begin{botherref}
\oauthor{\bsnm{Freedman}, \binits{D.}},
\oauthor{\bsnm{Diaconis}, \binits{P.}}:
On the histogram as a density estimator: {{L2}} theory
\textbf{57}(4),
453--476
\doiurl{10.1007/BF01025868}
\end{botherref}
\endbibitem

\bibitem[\protect\citeauthoryear{Patrone and Rosch}{}]{patrone_beyond_2017}
\begin{botherref}
\oauthor{\bsnm{Patrone}, \binits{P.N.}},
\oauthor{\bsnm{Rosch}, \binits{T.W.}}:
Beyond histograms: {{Efficiently}} estimating radial distribution functions via
  spectral {{Monte Carlo}}
\textbf{146}(9),
094107
\doiurl{10.1063/1.4977516}
\end{botherref}
\endbibitem

\bibitem[\protect\citeauthoryear{Hawat et~al.}{}]{hawat_estimating_2023}
\begin{botherref}
\oauthor{\bsnm{Hawat}, \binits{D.}},
\oauthor{\bsnm{Gautier}, \binits{G.}},
\oauthor{\bsnm{Bardenet}, \binits{R.}},
\oauthor{\bsnm{Lachièze-Rey}, \binits{R.}}:
On estimating the structure factor of a point process, with applications to
  hyperuniformity
\textbf{33}(3),
61
\doiurl{10.1007/s11222-023-10219-1}
\end{botherref}
\endbibitem

\end{thebibliography}
\end{document}